%% file: main.tex
\documentclass[acmtog]{acmart}
\acmSubmissionID{265}

\usepackage[utf8]{inputenc}
\usepackage{booktabs} % For formal tables
\usepackage[ruled]{algorithm2e} % For algorithms
\usepackage{tabularx}
\usepackage{multirow}
\usepackage{subcaption}
\usepackage{textcomp}
\usepackage{color}
\usepackage{colortbl}
\usepackage{mathtools}
\usepackage{siunitx}
\usepackage{makecell}
\usepackage[usestackEOL]{stackengine}

\input{macro}

\setcitestyle{square,nosort}

\SetAlFnt{\small}
\SetAlCapFnt{\small}
\SetAlCapNameFnt{\small}
\SetAlCapHSkip{0pt}
\IncMargin{-\parindent}

\setcopyright{acmlicensed}\acmJournal{TOG}
\acmYear{2020}\acmVolume{39}\acmNumber{6}\acmArticle{260}\acmMonth{12} \acmDOI{10.1145/3414685.3417821}

\newcommand{\ignore}[1]{}
\begin{document}

% \title{Light Stage Upsampling: Towards Continuous High-Frequency Relighting with Sparse Samples}

\title{Light Stage Super-Resolution: Continuous High-Frequency Relighting}

\author{Tiancheng Sun}
\email{tis037@cs.ucsd.edu}
\author{Zexiang Xu}
\email{zexiangxu@cs.ucsd.edu}
\affiliation{\institution{University of California, San Diego}}
\author{Xiuming Zhang}
\email{xiuming@csail.mit.edu}
\affiliation{\institution{Massachusetts Institute of Technology}}
\author{Sean Fanello}
\email{seanfa@google.com}
\author{Christoph Rhemann}
\email{crhemann@google.com}
\author{Paul Debevec}
\email{debevec@google.com}
\affiliation{\institution{Google}}
\author{Yun-Ta Tsai}
\email{yuntatsai@google.com}
\author{Jonathan T. Barron}
\email{barron@google.com}
\affiliation{\institution{Google Research}}
\author{Ravi Ramamoorthi}
\email{ravir@cs.ucsd.edu}
\affiliation{\institution{University of California, San Diego}}

\input{0-abstract}

\begin{CCSXML}
<ccs2012>
<concept>
<concept_id>10010147.10010371.10010382.10010385</concept_id>
<concept_desc>Computing methodologies~Image-based rendering</concept_desc>
<concept_significance>500</concept_significance>
</concept>
<concept>
<concept_id>10010147.10010257.10010293.10010294</concept_id>
<concept_desc>Computing methodologies~Neural networks</concept_desc>
<concept_significance>500</concept_significance>
</concept>
</ccs2012>
\end{CCSXML}

\ccsdesc[500]{Computing methodologies~Image-based rendering}
\ccsdesc[500]{Computing methodologies~Neural networks}

\keywords{Portrait relighting, Image-based relighting.}

\newcommand{\teaserwidth}{1.65in}
\begin{teaserfigure}
\begin{tabular}{@{}c@{\quad}c@{\quad}c@{}}
    \begin{subfigure}[b]{.48\textwidth}
    \centering
    \includegraphics[width=\teaserwidth]{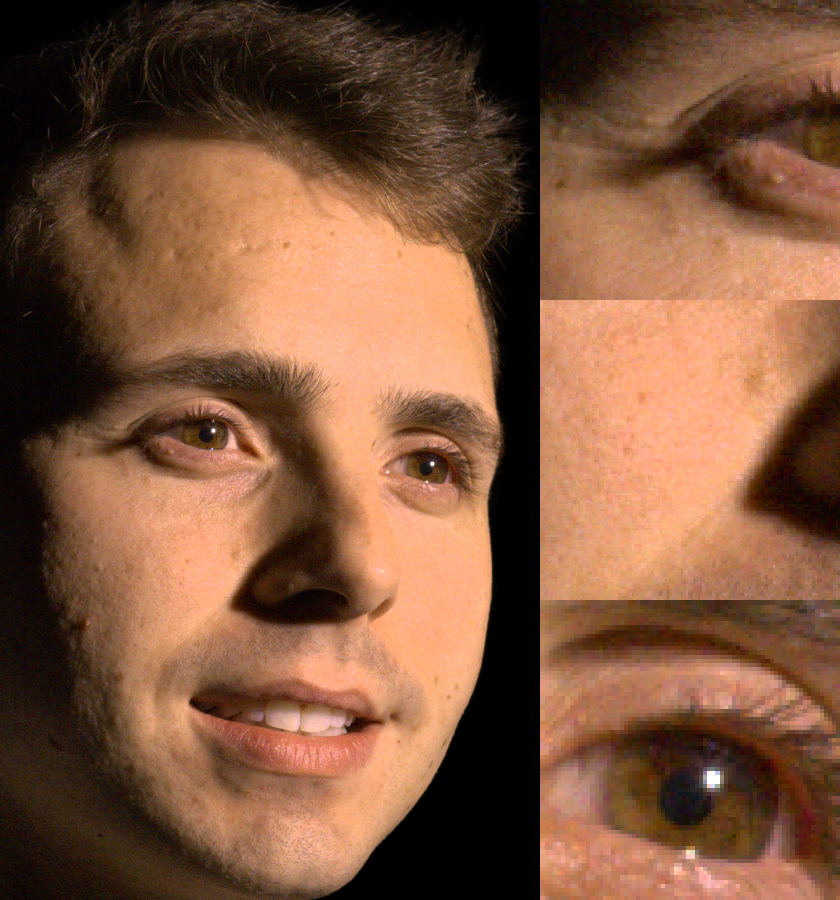}
    \includegraphics[width=\teaserwidth]{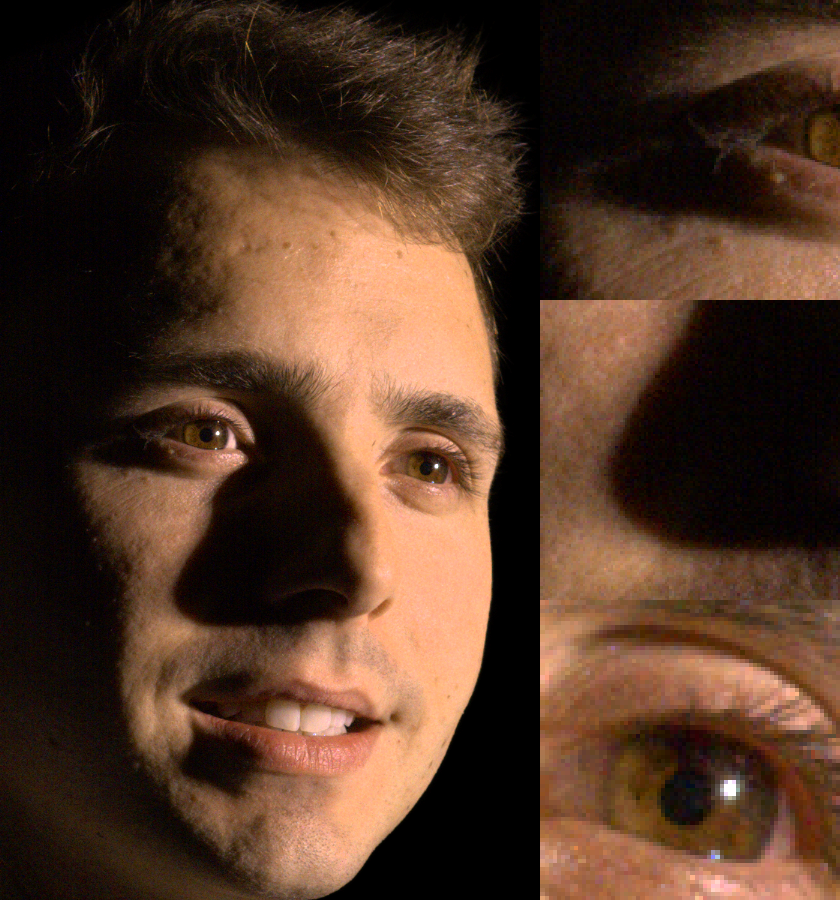}
    \caption{$\image_1$ and $\image_2$, captured light stage images for adjacent lights $\light_1$ and $\light_2$}\label{subfig:teaser_input}
    \end{subfigure}
    &
    \begin{subfigure}[b]{.24\textwidth}
    \centering
    \includegraphics[width=\teaserwidth]{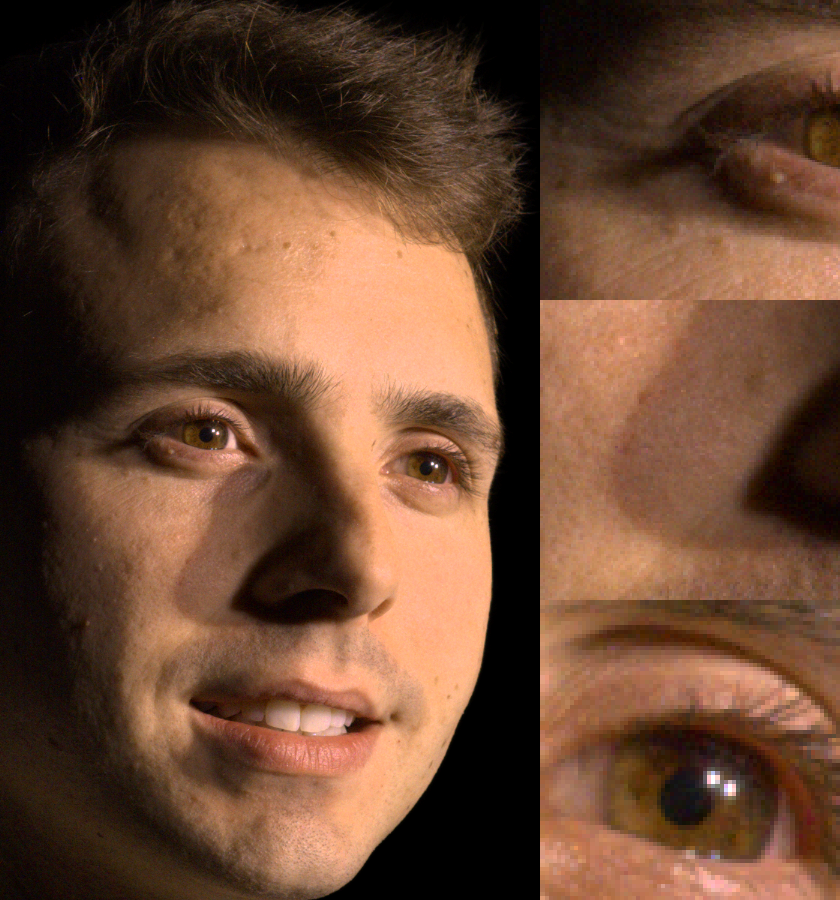}
    \caption{Blended images $(\image_1 + \image_2)/2$}\label{subfig:teaser_linear}
    \end{subfigure}
    &
    \begin{subfigure}[b]{.24\textwidth}
    \centering
    \includegraphics[width=\teaserwidth]{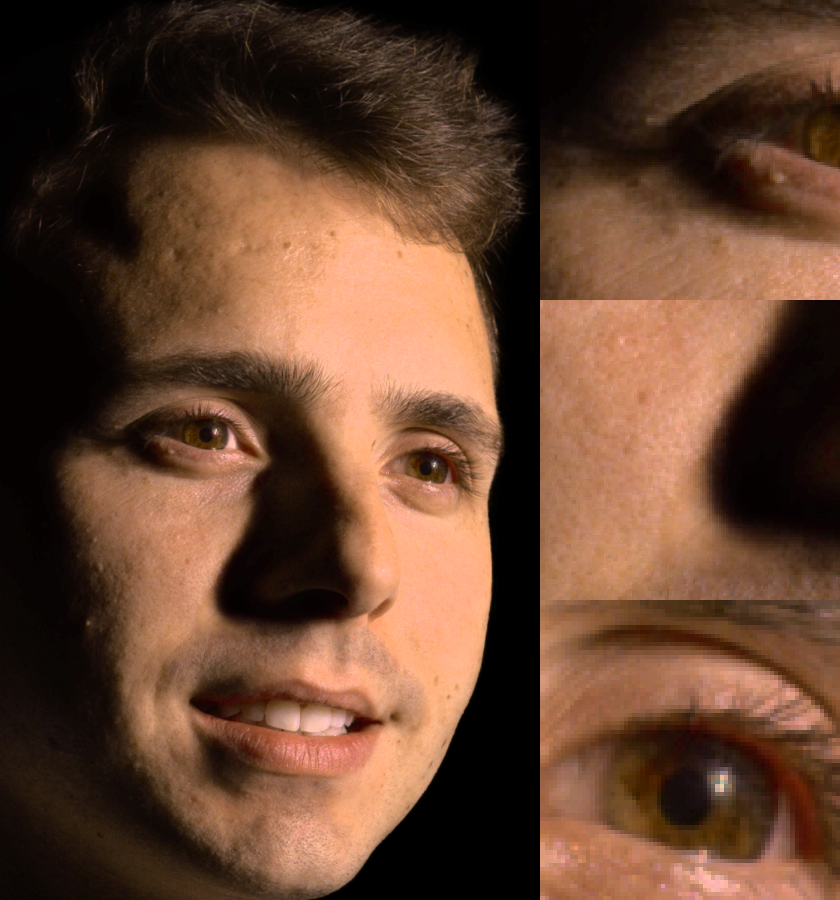}
    \caption{Our rendering $\predimage((\light_1 + \light_2)/2 )$}\label{subfig:teaser_ours}
    \end{subfigure}
    \end{tabular}
    % \vspace*{-.2in}
    \caption{
    Though the light stage is a powerful tool for relighting human subjects, its renderings suffer because adjacent lights of the stage are separated by some distance (a).
    Using conventional image blending techniques to reconstruct the image corresponding to a ``virtual'' light that lies between the stage's actual lights therefore results in ghosting in shadowed and specular regions (b), seen here on the subject's eyes and cheek.
    By training a deep neural network to regress from a light direction to an image, our model is able to synthesize accurate renderings of the subject under arbitrary virtual light directions --- as the light moves, highlights and shadows move smoothly instead of incorrectly blending together, thereby enabling realistic high-frequency relighting effects (c).
    These images have been manually but uniformly brightened and color-corrected, and are rendered with insets to highlight detail.
    % \barron{This figure needs to demonstrate the problem that we're solving, which is hard to convey without an illustration. A proposal: First we show (a,b) two adjacent OLAT images, and (c) their sum, with zoomed in crops on a cast shadow and a right specular highlight. This gets across the problem. Then we should (d), the output of our model for the point halfway in between the two OLAT images, where you can see that the shadow and highlight *move* but don't ghost, and maybe (e) the sum of the outputs of the model for all points in between the two OLAT images, which produces a soft shadow and a nice line of specular highlights, to contrast with (c). This maybe should be two separate figures, one where we show a final rendering accord to an environment map, and another where we motivate the problem and show our actual solution to that problem.}
    }
    \label{fig:teaser}
\end{teaserfigure}

\maketitle
\renewcommand{\shortauthors}{Sun \etal}

\input{1-intro}

\input{2-relatedwork}

\input{3-model}
\input{4-eval}

\input{5-app}
\input{6-conclude}

\begin{acks}
This work was supported in part by NSF grants 1617234, 1703957 ONR grants N000141712687 and N000142012529, a Google Fellowship, the Ronald L. Graham Chair, and the UC San Diego Center for Visual Computing. Thanks to anonymous reviewers for the valuable feedback.
\end{acks}

% Bibliography
\bibliographystyle{ACM-Reference-Format}
\bibliography{ref,combined}

\end{document}

%% file: macro.tex
\usepackage{xcolor}
\usepackage{textcomp}
\usepackage{xspace}

\definecolor{gray}{rgb}{0.5,0.5,0.5}
\definecolor{purple}{rgb}{0.7,0.3,0.7}
\definecolor{blue}{rgb}{0,0,1}
\definecolor{darkblue}{rgb}{0,0,0.6}
\definecolor{orange}{rgb}{1,.5,0} % something readable but different from todo
\definecolor{red}{rgb}{1,0,0} % something readable but different from todo

\newcommand{\changed}[1]{#1}

\newcommand{\Mat}[1]{{\ensuremath{\mathbf{\uppercase{#1}}}}} %Matrix 
\DeclarePairedDelimiterX{\norm}[1]{\lVert}{\rVert}{#1}
\DeclarePairedDelimiterX{\abs}[1]{\lvert}{\rvert}{#1}

\newcommand{\mask}{\Mat{M}}

\newcommand{\image}{\Mat{I}}
\newcommand{\light}{{\boldsymbol\ell}}
\newcommand{\predimage}{\image}
\newcommand{\querylight}{\light}

\newcommand{\activeset}{\mathrm{A}}
\newcommand{\weight}{\mathrm{W}}

\newcommand{\network}{\mathrm{\Phi}}
\newcommand{\relight}{\network}
\newcommand{\encoder}{\network_{e}}
\newcommand{\decoder}{\network_{d}}

\newcommand{\lightcount}{n}

\newcommand{\lossfun}{\mathcal{L}}

\definecolor{MyDarkBlue}{rgb}{0,0.08,1}
\definecolor{MyDarkGreen}{rgb}{0.02,0.6,0.02}
\definecolor{MyDarkRed}{rgb}{0.8,0.02,0.02}
\definecolor{MyDarkOrange}{rgb}{0.70,0.35,0.02}
\definecolor{MyPurple}{RGB}{0.43,0,1.}
\definecolor{MyRed}{rgb}{1.0,0.0,0.0}
\definecolor{MyGold}{rgb}{0.75,0.6,0.12}
\definecolor{MyDarkgray}{rgb}{0.66, 0.66, 0.66}

\newcommand{\barron}[1]{\textcolor{orange}{[Barron: #1]}}
\newcommand{\sean}[1]{\textcolor{MyPurple}{[Sean: #1]}}

\newcommand{\etal}{\textit{et al}. }
\newcommand{\ie}{\textit{i}.\textit{e}., }

\newcommand{\etc}{\textit{etc}. }

%% file: 0-abstract.tex
\begin{abstract}

The light stage has been widely used in computer graphics for the past two decades, primarily to enable the relighting of human faces. By capturing the appearance of the \changed{human} subject under different light sources, one obtains the light transport matrix of that subject, which enables image-based relighting in novel environments. However, due to the finite number of lights in the stage, the light transport matrix only represents a sparse sampling on the entire sphere. As a consequence, relighting the subject with a point light or a directional source that does not coincide exactly with one of the lights in the stage requires interpolation and resampling the images corresponding to nearby lights, and this leads to ghosting shadows, aliased specularities, and other artifacts. 
To ameliorate these artifacts and produce better results under arbitrary high-frequency lighting, this paper proposes a learning-based solution for the ``super-resolution'' of scans \changed{of human faces} taken from a light stage. Given an arbitrary ``query'' light direction, our method aggregates the captured images corresponding to neighboring lights in the stage, and uses a neural network to synthesize a rendering of the \changed{face} that appears to be illuminated by a ``virtual'' light source at the query location.
This neural network must circumvent the inherent aliasing and regularity of the light stage data that was used for training, which we accomplish through the use of regularized traditional interpolation methods within our network.
Our learned model is able to produce renderings for arbitrary light directions that exhibit realistic shadows and specular highlights, and is able to generalize across a wide variety of subjects.
Our super-resolution approach enables more accurate renderings of human subjects under detailed environment maps, or the construction of simpler light stages that contain fewer light sources while still yielding comparable quality renderings as light stages with more densely sampled lights.
%Ravi: Commented out for now, add back as needed.
%\tc{change the words depending on the results}
%\tc{Decide how far shall we talk about generalization after paper is complete.}
% Our results also point toward more general solutions for image-based relighting and light transport acquisition with many fewer light
% Even though we focused on the very specific task of light stage upsampling for faces, it moves towards a general solution for lighting interpolation in light transport acquisition, enabling much sparser sets of lighting to be used for image-based relighting, while preserving high-frequency effects. 

%\xiuming{light stage has been widely used for reflectance capturing and image-based rendering in the graphics community. }

\end{abstract}

%% file: 1-intro.tex
\section{Introduction}

A central problem in computer graphics and computer vision is that of acquiring some observations of an object, and then producing photorealistic relit renderings of that object.
% Such renderings of human faces are of particular interest, both due to their many practical uses within consumer photography and the special effects industry, but also because they serve as a particularly challenging example due to their complexity and \changed{human's high sensitivity to facial appearance}.
Of particular interest are renderings of human faces, which have many practical uses within consumer photography and the visual effects industry, but also serve as a particularly challenging case due to their complexity and the high sensitivity of the human visual system to facial appearance.
A light stage represents an effective solution for this task: by programmatically activating and deactivating several LED lights arranged in a sphere while capturing synchronized images, the light stage acquires a full reflectance field for a \changed{human} subject, which we refer to as a ``one-light-at-a-time'' (OLAT) image set.
Because light is additive, this OLAT scan represents a lighting ``basis'', and the subject can be relit according to some desired environment map by simply projecting that environment map onto the light stage basis~\cite{debevec2000acquiring}. 

Though straightforward and theoretically elegant, this classic relighting approach has a critical limitation. The lights on the light stage are usually designed to be small and distant from the subject, so that they are well-approximated as directional light sources. As a consequence, realistic high-frequency effects such as sharp cast shadows and specular highlights are present in the captured OLAT images. 
In order to achieve photorealistic relighting results under \emph{all} possible lighting conditions, the lights must be placed closely enough on the sphere of the stage such that shadows and specularities in the captured images of adjacent lights ``move'' by less than one pixel.
However, practical constraints (the cost and size of each light, and the difficulty of powering and synchronizing many lights) discourage the construction of light stages with very high densities of lights. Even if such a high-density light stage could be built, the time to acquire an OLAT increases linearly with the number of lights, and this makes human subjects (which must be stationary during OLAT acquisition) difficult to capture. For these reasons, even the most sophisticated light stages in existence today contain only a few hundred lights that are spaced many degrees apart.
This means that the OLAT scans from a light stage are \emph{undersampled} with respect to the angular sampling of lights, and the rendered images using conventional approaches will likely contain \emph{ghosting}.  Attempting to render an image using a ``virtual'' light source that lies in between the real lights of the stage by applying a weighted average on adjacent OLAT images will not produce a soft shadow or a streaking specularity, but will instead produce the superposition of multiple sharp shadows and specular dots (see Fig.~\ref{fig:teaser}b). 

This problem can be mitigated by imaging subjects that only exhibit low-frequency reflectance variation, or by performing relighting using only low-frequency environment maps. However, most human subjects have complicated material properties (specularities, scattering, \etc) and real-world environment maps frequently exhibit high-frequency variation (bright light sources at arbitrary locations), which often results in noticeable artifacts as shown in Fig.~\ref{fig:teaser}b.
To this end, we propose a learning based solution for super-resolving the angular resolution of light stage scans \changed{of human faces}. Given an OLAT scan \changed{of a human face} with finite images and the direction of a desired ``virtual'' light, our model predicts a complete high-resolution RGB image that appears to have been lit by a light source from that direction, even though that light is not present in our light stage (see Fig.~\ref{fig:teaser}c). 
Our robust solution for ``upsampling'' the number of lights, which we refer to as {\em light stage super-resolution}, can additionally enable the construction of simpler light stages with fewer lights, thereby reducing cost and increasing the frame rate at which subjects can be scanned.
Our algorithm can also produce better rendered images for applications that require light stage data for training, such as portrait relighting or shadow removal. Casual users can then utilize these algorithms on a single cellphone without requiring capture inside a light stage. \changed{Note that we focus only on human face relighting within a light stage. While we believe the methods herein could be applied more broadly, a comprehensive system for general object relighting remains a topic of future work.}

Our algorithm (Sec.~\ref{sec:model}) must work with the inherent aliasing and regularity of the light stage data used for training. We address this by combining the power of deep neural networks with the efficiency and generality of conventional linear interpolation methods. Specifically, we use an active set of closest lights within our network (Sec.~\ref{sec:activeset}) and develop a novel alias-free pooling approach to combine their network activations (Sec.~\ref{sec:pooling}) using a weighting operator guaranteed to be smooth when lights enter or exit the active set. 
Our network allows us to \emph{super-resolve} an OLAT scan \changed{of a human face}: we can take our learned model and repeatedly query it with thousands of light directions, and treat the resulting set of synthesized images as though they were acquired by a physically-unconstrained light stage with an unbounded sampling density. As we will demonstrate, these super-resolved ``virtual'' OLAT scans allow us to produce photorealistic renderings of \changed{human faces} with arbitrarily high-frequency illumination content.

%% file: 2-relatedwork.tex
\section{Related Work}

The angular undersampling from the light stage relates to much work over the past two decades on a frequency analysis of light transport~\cite{invrend,Imari,Fredo}, and can also be related to analyses of sampling rate in image-based rendering~\cite{plenoptic} for the related problem of view synthesis~\cite{llff}. This problem also bears some similarities to multi-image super-resolution~\cite{milanfar2010super} and angular super-resolution in the light field~\cite{kalantari2016learning,Cheng_2019_CVPR_Workshops}, where aliased observations are combined to produce interpolated results. In this paper, we leverage priors and deep learning to go beyond these sampling limits, upsampling or super-resolving a sparse input light sampling on the light stage to achieve continuous high-frequency relighting.  

% Our problem bears some similarities to the classic problem of multi-image super-resolution---the task of taking multiple (shifted) input images of a subject and fusing them together to produce a higher resolution output image.
% In contrast, our work attempts to perform angular (rather than spatial) super-resolution with respect to light direction. But this analogy offers some insight, as classic image super-resolution requires aliased observations to work well \cite{tsai1984multiframe}, and this criterion is satisfied in our context through the use of small LED lights within the light stage. The idea of angular super-resolution has been explored in light field super-resolution~\cite{Cheng_2019_CVPR_Workshops}, though there ``angular'' refers to the angle of incident light upon the camera sensor, while here it refers to the angle of incident light upon the subject being imaged.

Recently, many approaches for acquiring a sparse light transport matrix have been developed, including methods based on compressive sensing~\cite{peers2009compressive,Sen3}, kernel Nystrom~\cite{Tong}, optical computing~\cite{Otoole} and neural networks~\cite{ren2013global,Ren2015,kang2018efficient}. However, these methods are not designed for the light stage and are largely orthogonal to our approach.  They seek to acquire the transport matrix for a fixed light sampling resolution with a sparse set of patterns, while we seek to take this initial sampling resolution and upsample or super-resolve it to much higher-resolution lighting (and indeed enable continuous high-frequency relighting).  
Most recently,~\cite{xu2018deep} proposed a deep learning approach for image-based relighting from only five lighting directions, but cannot reproduce very accurate shadows.  While we do use many more lights, we achieve significantly higher-quality results with accurate shadows.  

The general approach of using light stages for image-based relighting stands in contrast to more model-based approaches. Traditionally, instead of super-resolving a light stage scan, one could use that scan as input to a photometric stereo algorithm~\cite{photometric_stereo}, and attempt to recover the normal and the albedo maps of the subject. More advanced techniques were developed to produce a parametric model of the geometry and reflectance for even highly specular objects~\cite{tunwattanapong2013acquiring}. There are also works that focus on recovering a parametric model from a single image~\cite{sfsnetSengupta18}, constructing a volumetric model for view synthesis~\cite{lombardi2018deep}, or even a neural representation of a scene~\cite{tewari2020state}. However, the complicated reflectance and geometry of human subjects is difficult to even parameterize analytically, let alone recover. Though recent progress may enable the accurate capture of human faces using parametric models, there are additional difficulties in capturing a complete portrait due to the complexity of human hair, eyes, ears, etc. Indeed, this complexity has motivated the use of image-based relighting via light stages in the visual effects industry for many years~\cite{tunwattanapong2011practical,debevec2012light}.  

%\tc{Modified this paragraph.}
Interpolating a reflectance function has also been investigated in the literature. \citet{masselus2004smooth} compare the errors of fitting the sampled reflectance function to various basis functions and conclude that multilevel B-Splines can preserve the most features. More recently, \citet{rainer2019neural} utilize neural networks to compress and interpolate sparsely sampled observations. However, these algorithms interpolate the reflectance function independently on each pixel and do not consider local information in neighboring pixels. Thus, their results are smooth and consistent in the light domain, but might not be consistent in the image domain.
\citet{fuchs2007superresolution} treat the problem as a light super-resolution problem, and is the most similar to our work. They use heuristics to decompose the captured images into diffuse and specular layers, and apply optical-flow and level-set algorithms to interpolate highlights and light visibility respectively. This approach works well on highly reflective objects, but as we will demonstrate, it usually fails on human skin which contains high frequency bumps and cannot be well modeled using only diffuse and specular terms.

\begin{figure*}
    \centering
    \includegraphics[width=\linewidth]{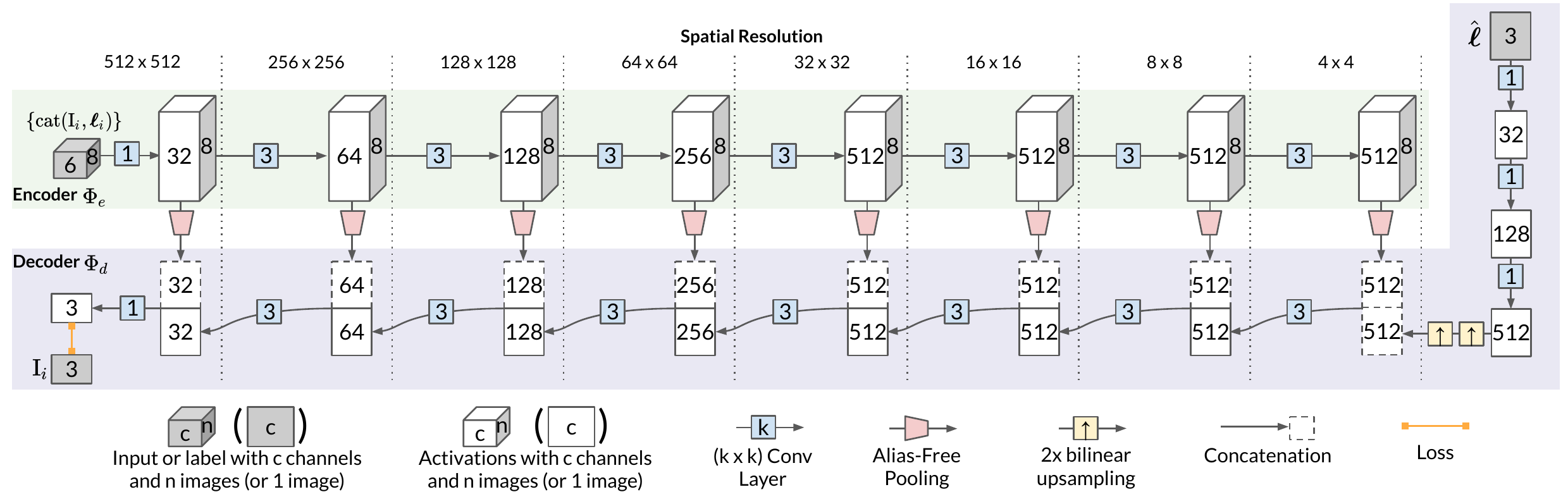}
    % \vspace*{-.25in}
    \caption{
    A visualization of our model architecture. The encoder of our model $\encoder(\cdot)$ takes as input a concatenation of the nearby OLAT images in the active set and their light directions, which are processed by a series of stride-2 conv layers. The resulting encoded activations of these 8 images at each level are then combined using the alias-free pooling described in Section~\ref{sec:pooling}, and skip-connected to the decoder. The decoder $\decoder(\cdot)$ takes as input the query light direction $\querylight$, processes it with fully connected layers and then upsamples it (along with the skip-connected encoder activations), and decodes the image using a series of stride-2 transposed conv layers.
    Whether or not a conv or transposed conv changes resolution is indicated by whether or not its edge spans two spatial scales.
    %The $\alpha_d$ parameters control the progressive training and growing of the network for each scale $d$ of the network by modulating the resolution at which input images are used and output images are compared to the ground truth, as described in Section~\ref{sec:loss}. 
    }
    % \vspace*{-.1in}
    \label{fig:architecture}
\end{figure*}

In recent years, light stages have also been demonstrated to be invaluable tools for generating training data for use in deep learning tasks~\cite{meka2019deep,guo_relightables:_2019,Sun2019,nestmeyer2019structural}.  This enables user-facing effects that do not require acquiring a complete light stage scan of the subject, such as ``portrait relighting'' from a single image~\cite{Sun2019, ApplePortraitMode} or VR experiences~\cite{guo_relightables:_2019}.
These learning-based applications suffer from the same undersampling issue as do conventional uses of light stage data.  For example, \citet{Sun2019} observe artifacts when relighting with environment maps that contain high-frequency illumination.  We believe our method can provide better training data and significantly improve many of these methods in the future.

%% file: 3-model.tex
\section{Model}
\label{sec:model}

An OLAT scan of a subject captured by a light stage consists of $n$ images, where each image is lit by a single light in the stage. The conventional way to relight the captured subject with an arbitrary light direction is to linearly blend the images captured under nearby lights in the OLAT scan. As shown in Fig.~\ref{fig:teaser}, this often results in ``ghosting'' artifacts on shadows and highlights. The goal of this work is to use machine learning instead of simple linear interpolation to produce higher-quality results.
Our model takes as input a query light direction $\querylight$ and a complete OLAT scan consisting of a set of paired images and light directions $\{ \image_i, \light_i \}$, and uses a deep neural network $\relight$ to obtain the predicted image $\predimage$,
\begin{equation}
    \predimage\left(\querylight\right) = \relight\left(\left\{ \image_i, \light_i \right\}_{i=1}^{\lightcount}, \querylight \right) .
    \label{equ:relight}
\end{equation}
This formalization is broad enough to describe some prior works on learning-based relighting~\cite{xu2018deep,meka2019deep}. While these methods usually operate by training a U-Net~\cite{ronneberger2015u} to map from a {\em sparse} set of input images to an output image, we focus on producing as high-quality as possible rendering results given the {\em complete} OLAT scan.
%However, directly applying this feed-forward approach to our problem domain is problematic. Because the input in our case is a complete OLAT scan, 
However, feeding all the captured images into a conventional CNN network is not tractable in terms of speed or memory requirements. In addition, this naive approach seems somewhat excessive for practical applications involving human faces. While complex translucency and interreflection may require multiple lights to reproduce, it is unlikely that \emph{all} images in the OLAT scan are necessary to reconstruct the image for any particular query light direction, especially given that barycentric interpolation requires only three nearby lights to produce a somewhat plausible rendering.
Our work attempts to find an effective and tractable compromise between these two extremes, in which the power of deep neural networks is combined with the efficiency and generality of nearest-neighbor approaches. This is accomplished by a linear blending approach that (like barycentric blending) ensures the output rendering is a smooth function of the input, where the blending is performed on the activations of a neural network's encoding of our input images instead of on the raw pixel intensities of the input images.

% The whole network structure is shown in Fig.~\ref{fig:architecture}. Formally speaking, an active set $\activeset(\querylight)$ is first constructed given the query light direction $\querylight$. Then, each OLAT image $\image_i$ and the corresponding light direction $\light_i$ from the active set are encoded into feature map sets $\feature_i$ by going through several convolutional layers:
% \begin{equation}\label{equ:encoder}
% \feature_i = \encoder\left(\image_i, \light_i\right), \textrm{where } i \in \activeset(\querylight).
% \end{equation}
% After that, we reduce these $k$ feature map sets into a single one by linearly blending them on the skip links using weight $\weight$. Together with the query light direction $\querylight$, the weighted feature map set is fed into the decoder, and produce the final prediction of the image $\predimage$:
% \begin{equation}\label{equ:decoder}
% \predimage\left(\querylight\right) = \decoder\left(\sum_{i\in\activeset(\querylight)}\weight(\querylight, \light_i)\feature_i, \querylight\right).
% \end{equation}
% The algorithm for choosing active set $\activeset$ and computing the weights $\weight$ for linear blending is detailed in the following section. 

Our complete network structure is shown in Fig.~\ref{fig:architecture}.
% Given the query light direction $\querylight$, we first construct an {\em active set} of $k$ neighboring OLAT image/light directions $\activeset(\querylight)$.
Given a query light direction $\querylight$, we identify the $k$ captured images in the OLAT scan whose corresponding light directions are nearby the query light direction, which we call \emph{active set} $\activeset(\querylight)$.
These OLAT images $\image_i$ and their corresponding light directions $\light_i$ are then each independently processed in parallel by the encoder $\encoder(\cdot)$ of our convolutional neural network (or equivalently, they are processed as a single ``batch''), thereby producing a multi-scale set of internal neural network activations that describe all $k$ images.
After that, the set of $k$ activations at each layer of the network are pooled into a single set of activations at each layer, which is performed using a weighted averaging where the weighting is a function of the query light and each input light $\weight(\querylight, \light_i)$. This weighted average is designed to remove the aliasing introduced by the nearest neighbor sampling in the active set selection stage.
% This collection of $k$ activations is then reduced down to a single set of multi-scale activations, which is then decoded via skip links by the second half of our network into the final synthesized image.
Together with the query light direction $\querylight$, these pooled feature maps are then fed into the decoder $\decoder(\cdot)$ by means of skip links from each level of the encoder, thereby producing the final predicted image $\predimage\left(\querylight\right)$.
Formally, our final image synthesis procedure is:
\begin{equation}
\predimage\left(\querylight\right) = \decoder\left(\sum_{i\in\activeset(\querylight)}\weight(\querylight, \light_i) \encoder\left(\image_i, \light_i\right), \querylight\right).
\label{equ:decoder}
\end{equation}
This hybrid approach of nearest-neighbor selection and neural network processing allows us to learn a single neural network that produces high quality results, and generalizes well across query light directions and across subjects in our OLAT dataset.% and requires significantly less compute and memory resources than alternative learned approaches \barron{Tiancheng: Confirm that these statements are true and backed up in the final draft, or delete this sentence}.

Our active set construction approach is explained in Section~\ref{sec:activeset}, our alias-free pooling is explained in Section~\ref{sec:pooling}, the network architecture is described in Section~\ref{sec:network}, and our progressive training procedure is discussed in Section~\ref{sec:loss}.

%\tc{Move from introduction}
%Designing a neural network architecture for light stage data super-resolution is not a straightforward task. We would like our model to generalize across light stage subjects, which means that our architecture must take as input images from an OLAT scan (as opposed to, perhaps, an architecture whose input is just a single virtual light location and whose weights encode the subject, that could be overfit to a single subject). But even the most capable modern hardware is not capable of training or evaluating a CNN that takes hundreds of high resolution images as input. Our architecture is designed to address these issues of generalization and efficiency: given a desired ``virtual'' light location, we take as input $k$ nearby neighbors in the OLAT scan (a small subset of the entire scan), we encode each of the neighboring images using a single CNN with shared weights, we pool the resulting $k$ sets of network activations, and we decode those averaged activations to produce our final image.

\begin{figure}
    \centering
    \includegraphics[width=\linewidth]{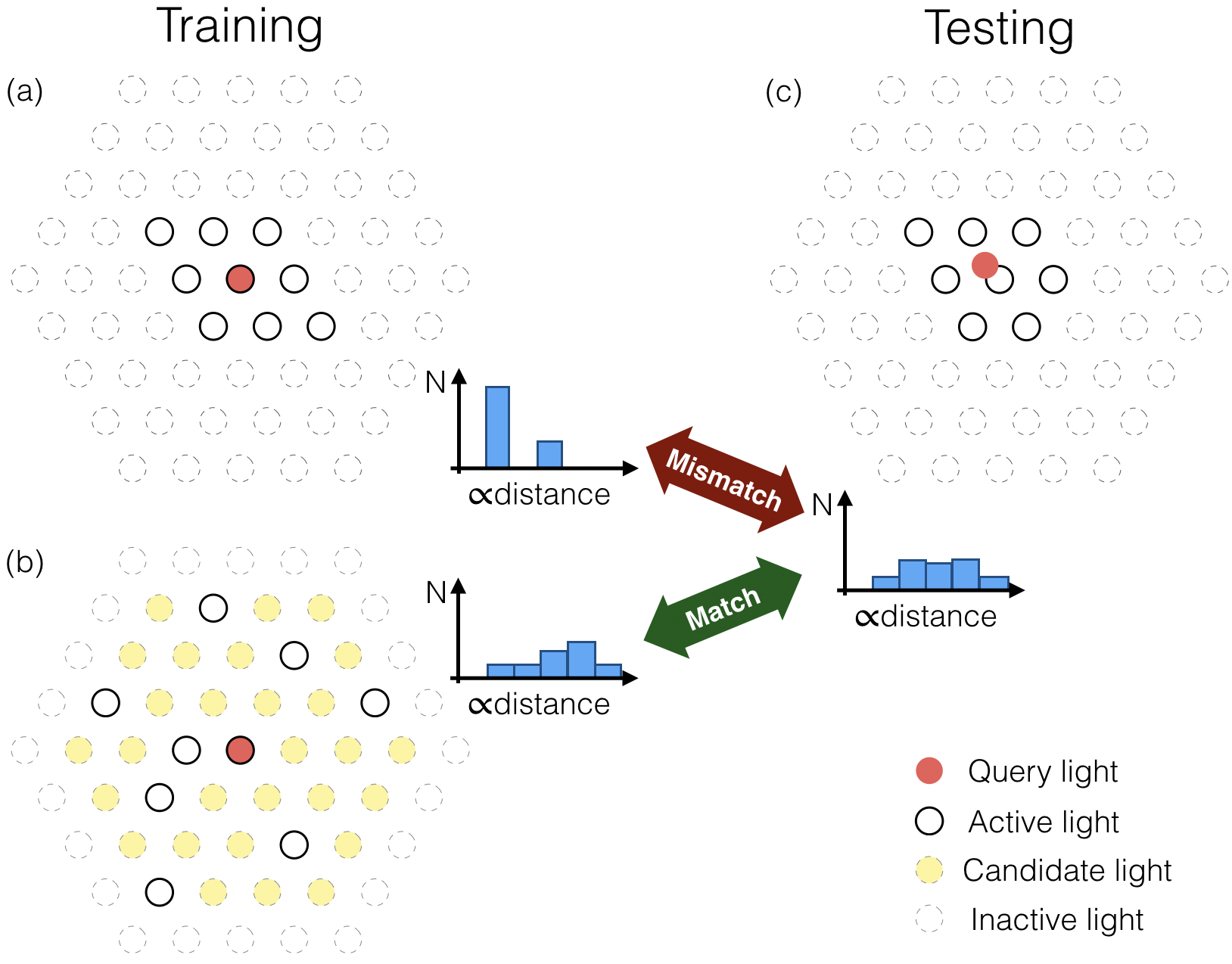}
    % \vspace*{-.2in}
    \caption{
    The OLAT images taken from a light stage have a uniform hexagonal pattern, which means that the distances between each light and its nearest neighbors is highly regular (a). In contrast, at test time we want to synthesize images corresponding to unseen light directions that do not lie on this hexagonal grid, and whose neighboring distances will therefore be irregular (c). During training we therefore sample a random subset of nearest neighbors for use in the active set of our model (b), which forces the network to adapt to challenging and irregular distributions of neighbor-distances that better match those that will be seen at test time.
    }
    \label{fig:activeset}
    % \vspace*{-.1in}
\end{figure}

\subsection{Active Set Selection}
\label{sec:activeset}

Light stages are conventionally constructed by placing lights on a regular hexagonal tessellation of a sphere (with some ``holes'' for cameras and other practical concerns), as shown in Fig.~\ref{fig:activeset}.
As discussed, at test time our model works by identifying the OLAT images and lights that are nearest to the desired query light direction, and averaging their neural activations.
But this natural approach, when combined with the regularity of the sampling of lights in the light stage, presents a number of problems for training our model.
First, we can only supervise our super-resolution model using ``virtual'' lights that exactly coincide with the real lights of the light stage, as these are the only light directions for which we have ground-truth images (this will also be a problem when evaluating our model, as will be discussed in Sec.~\ref{sec:eval}).
Second, this regular hexagonal sampling means that, for any given light in the stage, the distances between it and its neighbors will always exhibit a highly regular pattern (Fig.~\ref{fig:activeset}a). For example, the $6$ nearest neighbors of every point on a hexagonal tiling are guaranteed to have exactly the same distance to that point. In contrast, at test time we would like to be able to produce renderings for query light directions that correspond to arbitrary points on the sphere, and those points will likely have irregular distributions of neighboring lights (Fig.~\ref{fig:activeset}c). This represents a significant deviation between our training data and our test data, and as such we should expect poor generalization at test time if we were to naively train on highly-regular sets of nearest neighbors.
%but test on highly-irregular sets of neighbors.

%training our model: the OLAT scans that we use as input to the model are also used to supervise the training of the model, which means that our ground truth is just as undersampled as our input. That is, we would like to train our model to synthesize images that lie between our observed images, but no such images exist in our training data. As such, naively training our model results in a network that produces accurate renderings at the positions where we have training data, but produces ghosting and blending artifacts at all other points, completely defeating the purpose of this model.

To address this issue, we adopt a different technique for sampling neighbors for use in our active set than what is used during test time. For each training iteration, we first identify a larger set of $m$ nearest neighbors near the query light (which in this case is identical to one of the real lights in the stage), and among them randomly select only $k<m$ neighbors to use in the active set (in practice, we use $m=16$ and $k=8$). As shown in Fig.~\ref{fig:activeset}b, this results in irregular neighbor sampling patterns during training, which simulates our test-time scenario wherein the query light is at a variety of locations in between the real input light sources. 
This approach shares a similar motivation as that of dropout~\cite{srivastava2014dropout} in neural networks, in which network activations are randomly set to $0$ during training to prevent overfitting. Here we instead randomly remove input images, which also has the effect of preventing the model from overfitting to the hexagonal pattern of the light stage while training our network, by forcing it to operate on more varied inputs.
%and preventing it model from overfitting to the hexagonal pattern of the light stage. 
%In practice, we choose the $m=24$ nearest neighbors as candidates for the active set, and from them randomly sample $k=8$ lights into the active set. 
Notice that the query light itself is included in the candidate set, to reflect the fact that during test-time the ``virtual'' query light may be next to a real light source. As we will show in Sec.~\ref{sec:eval} and in the supplementary video, this active set selection approach results in a learned model whose synthesized shadows move more smoothly and at a more regular rate than is achieved with a naive nearest-neighbor sampling approach.

%\barron{Connect this to dropout , as it's doing something similar: instead of randomly dropping out activations within a network we're randomly dropping out input tensors, and this also has the effect of regularizing our network to operate on more varied inputs --- just as dropout prevents overfitting, dropping inputs in this way prevents our model from overfitting to the input image sampling pattern of the light stage. Dropout is also motivated as a way to perform model averaging, and this is much more explicitly what we're doing here.}.

% \barron{there's a tension between setting $m$ to be very large, where some neighbors are so distant that they're very unlike what we'd see at test time, and setting $m$ to be very small, where the sampling pattern is too regular.}

\begin{figure}
    \centering
    \includegraphics[width=\linewidth]{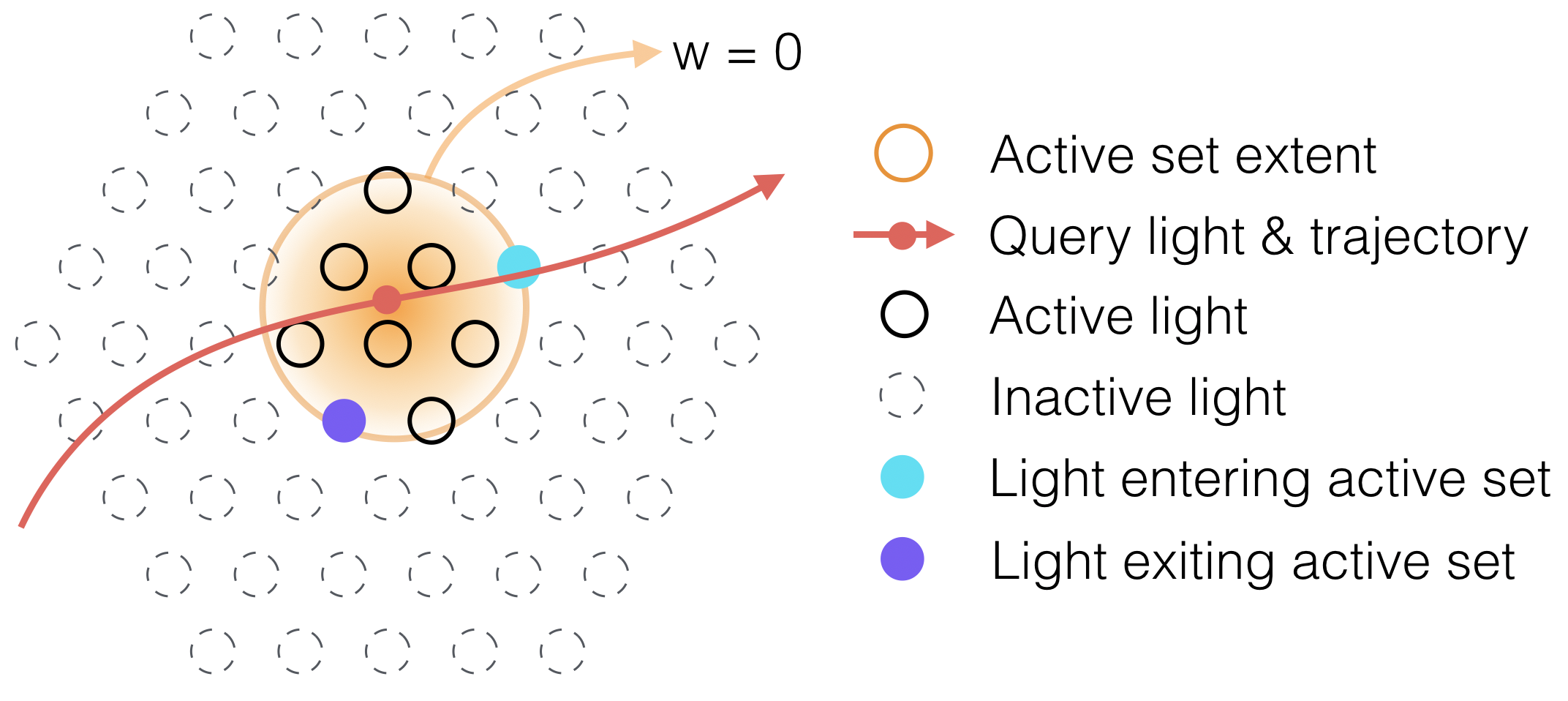}
    % \vspace*{-.2in}
    \caption{Varying the query light direction will cause OLAT images to leave and enter the active set of our model, which introduces aliasing that, if unaddressed, results in jarring temporal artifacts in our renderings. To address this, we use an ``alias-free pooling'' technique to ensure that the network activations of each OLAT image are averaged in a way that suppresses this aliasing. We use a weighted average where the weights are smooth, and are exactly zero at the point where lights enter and leave the active set.
    }
    \label{fig:weight}
    % \vspace{-.1in}
\end{figure}

\subsection{Alias-Free Pooling}
\label{sec:pooling}

A critical component in our model is the design of the skip links from each level of the encoder of our model to its corresponding level in the decoder. This model component is responsible for the network activations corresponding to the 8 images in our active set and reducing them to one set of activations corresponding to a single output, which will then be decoded into an image. This requires a pooling operator for these 8 images. This pooling operator must be permutation-invariant, as the images in our active set may correspond to any OLAT light direction and may be presented in any order. Standard permutation-invariant pooling operators, such as average-pooling or max-pooling, are not sufficient for our case, because they do not suppress \emph{aliasing}.
As the query light direction moves across the sphere, images will enter and leave the active set of our model, which will cause the network activations within our encoder to change suddenly (see Fig.~\ref{fig:weight}). If we use simple average-pooling or max-pooling, the activations in our decoder will also vary abruptly, resulting in unrealistic flickering artifacts or temporal instability in our output renderings as the light direction varies. \changed{In other words, the point sampled signal should go through an effective prefiltering process in order to suppess the artifacts.}

The root cause of this problem is that our active set is an aliased observation of the input images, and average- or max-pooling allows this aliasing to persist. We therefore introduce a technique for alias-free pooling to address this issue. We use a weighted average as our pooling operator where the weight of each item in our active set is a continuous function of the query light direction, and where the weight of each item is guaranteed to be zero at the moment it enters or leaves the active set.
We define our weighting function between the query light direction $\querylight$ and each OLAT light direction $\light_i$ as follows:
\newcommand{\scalefactor}{s}
\newcommand{\rawweight}{\widetilde{\weight}}  % Because tildes often indicate "lifted" variables.
\begin{equation}
\begin{gathered}
    \rawweight(\querylight, \light_i) = \operatorname{max}\left(0,  e^{s\left(\querylight\cdot\light_i - 1\right)} - \min_{j \in \activeset(\querylight)} e^{s\left(\querylight\cdot\light_j - 1\right)}\right), \\
    \weight(\querylight, \light_i) = \frac{\rawweight(\querylight, \light_i)}{\sum_j \rawweight(\querylight, \light_j)},
    \label{equ:weight}
\end{gathered}
\end{equation}
where $\scalefactor$ is a learnable parameter that adjusts the decay of the weight with respect to the distance and each $\light$ is a normalized vector in 3D space. During training, parameter $\scalefactor$ will be automatically adjusted to balance between selecting the nearest neighbor ($\scalefactor = +\infty$) and taking an unweighted average of all neighbors ($\scalefactor = 0$).

Our weighting function is an offset spherical Gaussian, similar to the normalized Gaussian distance between the query light's Cartesian coordinates and those of the other lights in our active set, but where we have subtracted out the unnormalized weight corresponding to the most distant light in the active set (and clipped the resulting weights at $0$).
This adaptive truncation is necessary because the lights in the light stage may be spaced irregularly (due to holes in the stage for cameras or other reasons),
which means that a fixed truncation may be too aggressive in setting weights to zero in regions where lights are sampled less frequently. We instead leverage the fact that %(because the number of lights in the active set is fixed) 
when a light exits the active set, a new light will enter it at exactly the same time with exactly the same distance to the query light. This allows us to truncate our Gaussian weights using the maximum distance in the active set, which ensures that lights have a weight of zero as they leave or enter the active set. This results in renderings that change smoothly as we move the query light direction.
%\ravi{Maybe some discussion of this being only $C^0$ continuous, but that's ok?}
%\barron{Yeah I think it would be easier to explain (and almost identical) if we used a shifted smoothstep function, which would be $C^0$ and $C^1$ continuous. Something we can do for the camera ready if we want.}

\subsection{Network Architecture}
\label{sec:network}

The remaining components of our model consist of the conventional building blocks used in constructing convolutional neural networks, and can be seen in Fig.~\ref{fig:architecture}.
The encoder of our network consists of $3\times3$ convolutional neural network blocks (with a stride of 2 so as to reduce resolution by half), each of which is followed by group normalization \cite{Wu_2018_ECCV} and a PReLU~\cite{he2015delving} activation function.
The number of hidden units for each layer begins at $32$ and doubles after each layer, but is clipped at $512$.
The input to our encoder is a set of $8$ RGB input images corresponding to the nearby OLAT images in our active set, each of which has been concatenated with the $xyz$ coordinate of its source light (tiled to every pixel) giving us 8 6-channel input images.

These images are processed along the ``batch'' dimension of our network, and so are treated identically at each level of the encoder.
These 8 images are then pooled down to a single ``image'' (\ie a single batch) of activations using the alias-free pooling approach of Section~\ref{sec:pooling}, each of which is concatenated onto the internal activations of the network's decoder.

The decoder of the network begins with a series of fully-connected (aka ``dense'') neural network blocks that take as input the query light direction $\querylight$, each of which is followed by instance normalization~\cite{ulyanov2016instance} and a PReLU activation function. These activations are then upsampled to $4 \times 4$ and used as the basis of our decoder. Each layer of the decoder consists of a $3 \times 3$ transposed convolutional neural network block (with a stride of 2 so as to double resolution) which is again followed by group normalization and a PReLU activation function. The input to each layer's conv block is a concatenation of the upsampled activations from the previous decoder level, with the pooled activations from the encoder that have been ``skip'' connected from the same spatial scale.
The final activation function before any output image is produced is a sigmoid function, as our images are normalized to $[0, 1]$.
Because our network is fully convolutional~\cite{long2015fully}, it can be evaluated on images of arbitrary resolution, with GPU memory being the only limiting factor.
We train on $512 \times 512$ resolution images for the sake of speed, and evaluate and test on $1024 \times 1024$ resolution images to maximize image quality. 

% Our network is fully convolutional~\cite{long2015fully}, and so is able to adapt to variable input image dimensions. We train on $512 \times 512$ resolution images for the sake of speed, and evaluate and test on $1024 \times 1024$ resolution images to maximize image quality. 
% \changed{Since our network is fully convolutional, note that we can test on images of arbitrary size, with GPU memory being the only bound on image resolution.}

\subsection{Loss Functions and Training Strategy}
\label{sec:loss}

We supervise the training of our model using an $L_1$ loss on pixel intensities. Formally, our loss function is:
\begin{equation}
    \lossfun_d = \sum_i\left\lVert \mask \odot \left(\image_i -  \predimage\left(\light_i\right)\right) \right\lVert_{1},
    \label{eq:lossfun}
\end{equation}
where $\image_i$ is the ground truth image under light $i$, and $\predimage\left(\light_i\right)$ is our prediction. When computing the loss over the image, we use a precomputed binary image \mask~to mask out pixels that are known to belong to the background of the subject.

During training, we construct each training data instance by randomly selecting a human subject in our training dataset and then randomly selecting one OLAT light direction $i$.
The image corresponding to that light $\image_i$ will be used as the ground-truth image our model will attempt to reconstruct, and the ``query'' light direction for our model will be the light corresponding to that image $\light_i$.
We then identify a set of 8 neighboring images/lights to include in our active set using the selection procedure described in Section~\ref{sec:activeset}.
Our only data augmentation is a randomly-positioned $512\times512$ crop of all images in each batch.

% \changed{Our network consists of a stack of convolutional operators and operates on high-resolution images. Although skip links can facilitate the gradient passing during training, it is still hard to train such a deep network. Thus, we use progressive strategy~\cite{karras2017progressive} to train our network. In general, we start by downsampled inputs to train only a part of our network, which is shallower and easier to train. Then, we gradually add in new conv layers and train higher resolution images at each stage, until we arrive at the original network and image resolution. In total, we train our network for 20,0000 iterations, using 8 NVIDIA V100 GPUs, which takes roughly 10 hours. Please see the detailed training procedure in Appendix~\ref{append:prog}.}

Progressive training has been found to be effective for accelerating and stabilizing the training of GANs for high-resolution image synthesis~\cite{karras2017progressive}, and though our model is not a GAN (but is instead a convolutional encoder-decoder architecture with skip connections) we found it to also benefit from a progressive training strategy.
We first inject downsampled image inputs directly into a coarse layer of our encoder and supervise training by imposing a reconstruction loss at a coarse layer of our decoder, resulting in a shallower model that is easier to train. As training proceeds, we add additional convolutional layers to the encoder and decoder, thereby gradually increasing the resolution of our model until we arrive at the complete network and the full image resolution. In total, we train our network for 200,000 iterations, using 8 NVIDIA V100 GPUs, which takes approximately 10 hours. Please see the detailed training procedure in the supplementary material.

Our model is implemented in Tensorflow~\cite{Tensorflow} and trained using Adam~\cite{KingmaB14} with a batch size of 1 (the ``batch'' dimension of our tensors is used to represent the $8$ images in our active set), a learning rate of $10^{-3}$, and default hyperparameter settings ($\beta_1 = 0.9, \beta_2 = 0.999, \epsilon=10^{-7}$). 

%\ravi{I found this (sub)section pretty hard to read.  I suppose this detail is needed, and you may not be able to do anything  about it, but if you can make it somehow easier to read.}

% Progressive training~\cite{karras2017progressive}
% \barron{todo: define $\alpha_d$ as a function of iteration (epoch?) $t$}
% Total number of steps = $T$.
% $K$ stages total.
% Linearly interpolate each $\upalpha_{\mathit{d}}$ as a function of $t$, such that only one $\upalpha_{\mathit{d}}$ is active at any times:
% \begin{equation}
%     \upalpha_{\mathit{d}} = \operatorname{max}\left(0, 1 - \left|\frac{K(T - t)}{T} - d\right|\right)
% \end{equation}
% We use $K=7$ stages and $T=$\todo{???} iterations \barron{that's the number of stages times the number of iterations per stage}

%% file: 4-eval.tex
\section{Evaluation}
\label{sec:eval}

\begin{figure*}
\begin{tabular}{@{}c@{\quad\,\,}c@{\,\,}c@{\,\,}c@{}}
    \begin{subfigure}[b]{.23\linewidth}
    \centering
    \includegraphics[width=\linewidth]{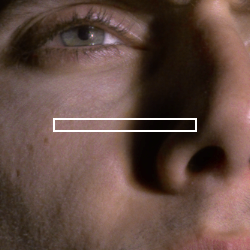}
    \caption{One rendering, for reference}\label{subfig:movingshadowa}
    \end{subfigure}
    &
    \begin{subfigure}[b]{.2562\linewidth}
    \centering
    \includegraphics[width=\linewidth]{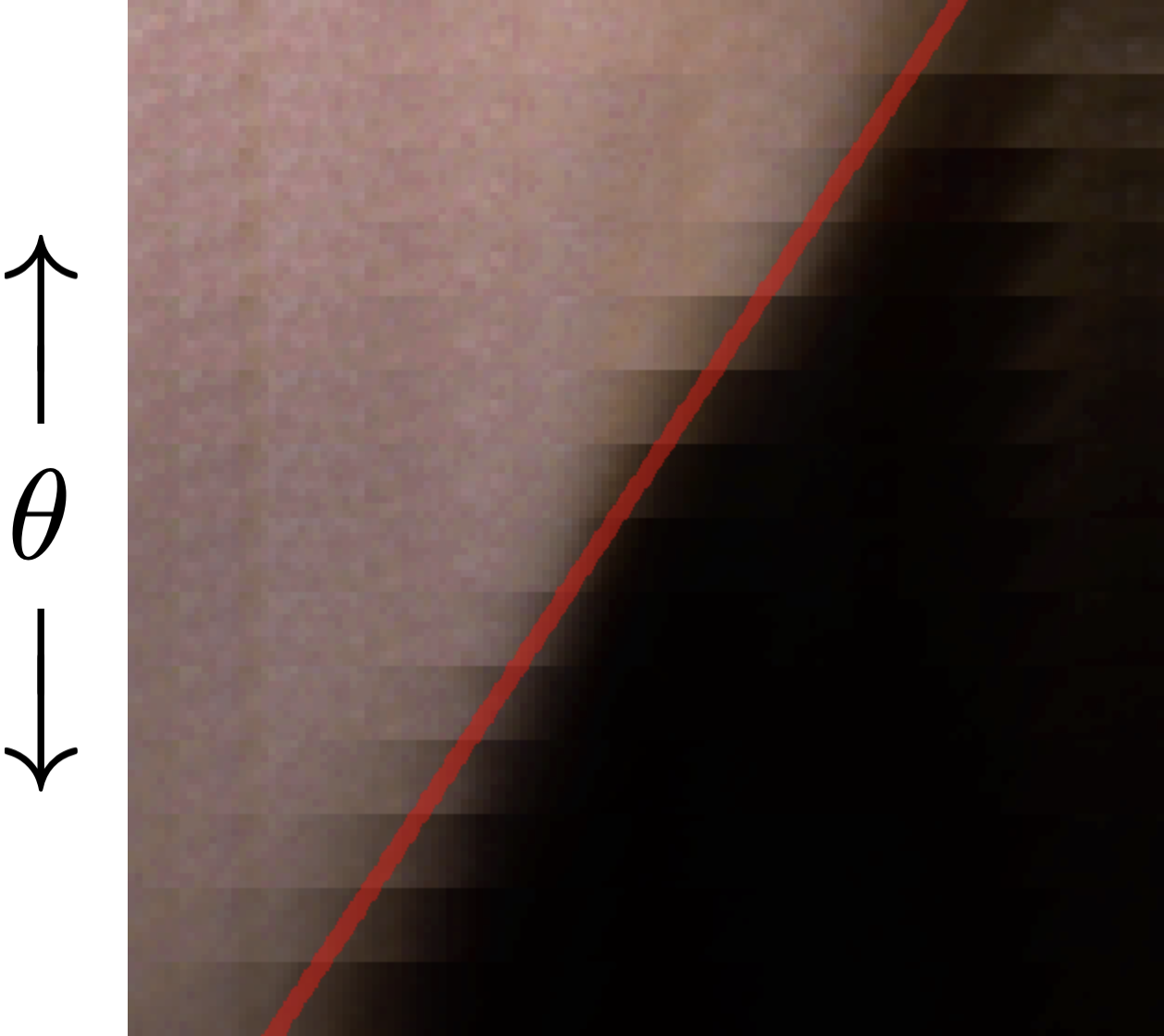}
    \caption{Our model}\label{subfig:movingshadowb}
    \end{subfigure}
    &
    \begin{subfigure}[b]{.23\linewidth}
    \centering
    \includegraphics[width=\linewidth]{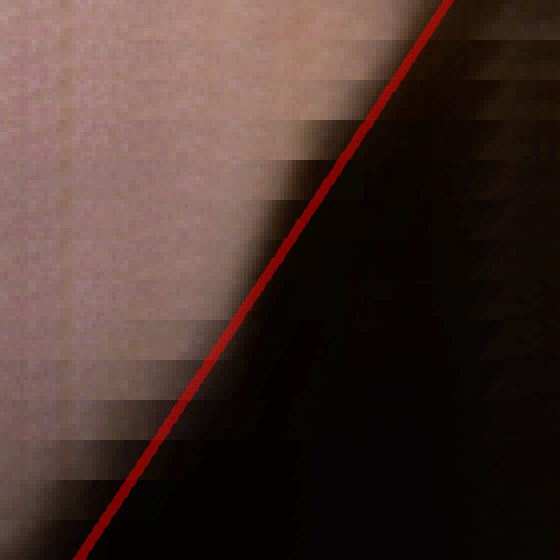}
    \caption{Our model w/ naive neighbors}\label{subfig:movingshadowc}
    \end{subfigure}
    &
    \begin{subfigure}[b]{.23\linewidth}
    \centering
    \includegraphics[width=\linewidth]{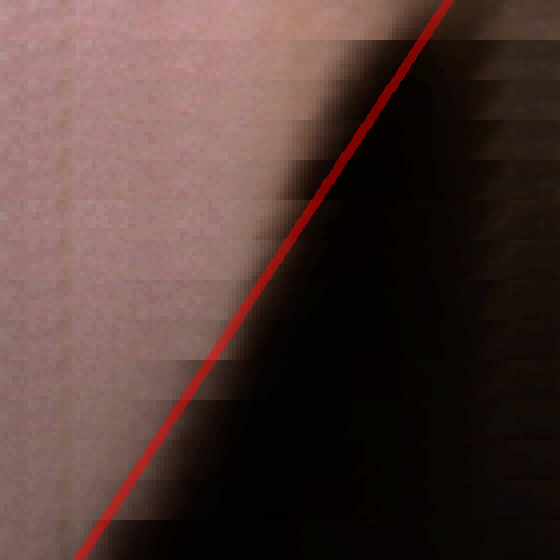}
    \caption{Our model w/ avg pooling}\label{subfig:movingshadowd}
    \end{subfigure}
\end{tabular}
% \vspace{-0.25cm}
\caption{
A visualization of how our learned model synthesizes renderings in which shadows move smoothly as a function of light direction.
In (a) we show a rendering from our model for some virtual light $\querylight$ with a horizontal angle of $\theta$, and highlight one image strip that includes a horizontal cast shadow.
In (b) we repeatedly query our model with $\theta$ values that should induce a linear horizontal translation of the shadow boundary in the image plane, and by stacking these image strips we can see this linear trend emerge (highlighted in red).
In (c) and (d) we do the same for ablations of our model that do not have our active-set random selection procedure nor our alias-free pooling, and we see that the resulting shadow boundary does not vary smoothly or linearly.
}
% \vspace{-0.3cm}
\label{fig:movingshadow}
\end{figure*}

We use the OLAT portrait dataset from~\cite{Sun2019}, which contains 22 subjects with multiple facial expressions captured using a light stage and a 7-camera system.
The light stage consists of 302 LEDs uniformly distributed on a spherical dome, and capturing a subject takes roughly 6 seconds. Each capture process produces an OLAT scan of a specific facial expression on each camera, which consists of 302 images, and we treat the OLAT scans from different cameras as independent OLAT scans. Because the subject is asked to stay still (and an optical flow algorithm~\cite{Wenger:2005:PRR} is applied to correct the small movements) the captured 302 images in each OLAT are aligned and only differ in lighting directions. We manually select 4 OLAT scans with a mixture of subjects and views for use as our validation set, and choose another 16 OLAT scans with good coverage of gender and diverse skin tones for use as training data. Our 16 training datasets only covers 5 of 7 cameras, and the remaining 2 are covered by the validation data. We train our network using all lights from our OLAT data in a canonical global lighting coordinate frame, which allows us to train a single network for all viewpoints in our training data. We train one single model for all subjects in our training dataset, which we found to match the performance of training an individual model for each subject.

% We train a single network for all viewpoints and subjects in our training data

% \changed{Our network is trained on global lighting coordinates with OLATs from different viewpoints, and does not require separate training for each viewpoint or subject.}

% \changed{The validation set includes unseen subjects and views from the training set}.
% \tc{Maybe talk a little bit on generalization}
%\sean{Maybe worth specifying you refer to 16 groundtruth olat, i.e. olats we are trying to predict :) I thought the model was trained just on 16 olats :D Is really enough to train on 16 to generalize to the whole dome?}

Empirically evaluating our model presents a significant challenge: our model is attempting to super-resolve an undersampled scan from a light stage, which means that the only ground-truth that is available for benchmarking is \emph{also} undersampled. In other words, the goal of our model is to accurately synthesize images that correspond to virtual lights in between the real lights of the stage --- but we do not have ground-truth images that correspond to those virtual lights. In addition, the model also needs to generalize to an unseen view and subject.
For these reasons, qualitative results (figures, videos) are preferred, and we encourage readers to view our figures and the accompanying video.
In the quantitative results presented here, we use held-out real images lit by real lights on our light stage as a validation set.
When evaluating one of these validation images, we do not use the active-set selection technique of Section~\ref{sec:activeset}, and instead just sample the $k=8$ nearest neighbors (excluding the validation image itself from the input). Holding out the validation image from the inputs is critical, as otherwise a model could simply reproduce the input image as an error-free output. This held-out validation approach is not ideal, as all such evaluations will follow the same regular sampling pattern of our light stage. This evaluation task is therefore more biased than the real task of predicting images away from the sampling pattern of the light stage.

Selecting an appropriate metric for measuring image reconstruction accuracy for our task is not straightforward.
Conventional image interpolation techniques often result in ghosting artifacts or duplicated highlights, which are perceptually salient but often not penalized heavily by traditional image metrics such as per-pixel RMSE. We therefore evaluate image quality using multiple image metrics: RMSE, the Sobolev $H^1$ norm~\cite{Ravi2}, DSSIM~\cite{SSIM}, and E-LPIPS~\cite{elpips}. RMSE measures pixel-wise error, the $H^1$ norm emphasizes image gradient error, while DSSIM and E-LPIPS approximate an overall perceptual difference between the predicted image and the ground truth. Still, images and videos are preferred for comparison.

\definecolor{Yellow}{rgb}{1,1, 0.6}
\definecolor{Orange}{rgb}{1,0.8, 0.6}
\definecolor{Red}{rgb}{1, 0.6, 0.6}
\begin{table}[t]
\caption{
Here we benchmark our model against prior work and ablations of our model on our validation dataset. We report the arithmetic mean of each metric across the validation set. The top three results of each metric are highlighted in red, orange, yellow, respectively. While ``Ours w/naive neighbors`` has the lowest error according to this evaluation, ``Our model`` performs better in our \emph{real} test-time scenario where the synthesized light does not lie in a regular hexagonal grid (see text and Fig.~\ref{fig:movingshadow} for details).
\label{table:compare}}
% \vspace{-.1in}
\begin{center}
\begin{tabular}{p{33mm}@{\,\,}|@{\,\,}c@{\quad}c@{\quad}c@{\quad}c}
Algorithm & RMSE & $H^1$ & DSSIM & E-LPIPS \\
\hline
Our model              & \cellcolor{Orange} 0.0160 & \cellcolor{Orange} 0.0203 &  \cellcolor{Orange} 0.0331 & \cellcolor{Orange} 0.00466  \\
Ours w/naive neighbors & \cellcolor{Red}  0.0156 &  \cellcolor{Red} 0.0199 & \cellcolor{Red} 0.0322 & \cellcolor{Red} 0.00449 \\
Ours w/avg-pooling     & 0.0203 &  0.0241 &  0.0413 & 0.00579 \\
\hline
Linear blending        & \cellcolor{Yellow} 0.0191 & \cellcolor{Yellow} 0.0232 & \cellcolor{Yellow} 0.0366 &  0.00503 \\
% \quad simple average & 0.0220 & 0.0258 & 0.0433 &  0.00613 \\
% \quad w/ alias-free weight & \cellcolor{Yellow} 0.0191 & \cellcolor{Yellow} 0.0232 & \cellcolor{Yellow} 0.0366 &  0.00503 \\
\citet{fuchs2007superresolution}          & 0.0195 & 0.0258 & 0.0382 & \cellcolor{Yellow} 0.00485\\
Photometric stereo     & 0.0284 & 0.0362 & 0.0968 & 0.00895\\
\citet{xu2018deep}     &&&& \\
\quad w/ 8 optimal lights    & 0.0410 & 0.0437 & 0.1262 & 0.01666\\
\quad w/ adaptive input     & 0.0259 & 0.0291 & 0.1156 & 0.00916\\
\citet{meka2019deep}   & 0.0505 & 0.0561 & 0.1308 & 0.01482\\
\end{tabular}
\end{center}
% \vspace{-.1in}
\end{table}

\setlength{\tabcolsep}{0pt} % General space between cols (6pt standard)
\newcommand{\fullwidth}{0.10\textwidth}
\newcommand{\cropwidth}{0.105\textwidth} 
\begin{figure*}
\centering
\begin{tabular}[c]{@{}c@{\,\,}c@{\,\,}c@{\,\,}c@{\,\,}c@{\,\,}c@{\,\,}c@{\,\,}c@{\,\,}c@{}}
{\small \Centerstack{(a) Ours\\(full image)}} & {\small (b) Groundtruth} & {\small (c) Ours} & {\small \Centerstack{(d) Linear\\blending}} & {\small \Centerstack{(e) Fuchs\\\etal~\shortcite{fuchs2007superresolution}}} & {\small \Centerstack{(f) Photometric\\stereo}} & {\small \Centerstack{(g) Xu\\ \etal~\shortcite{xu2018deep} w/\\optimal sample}} & {\small \Centerstack{(h) Xu\\\etal~\shortcite{xu2018deep} w/\\adaptive sample}} & {\small \Centerstack{(i) Meka\\\etal~\shortcite{meka2019deep}}}\\
\includegraphics[width=\cropwidth]{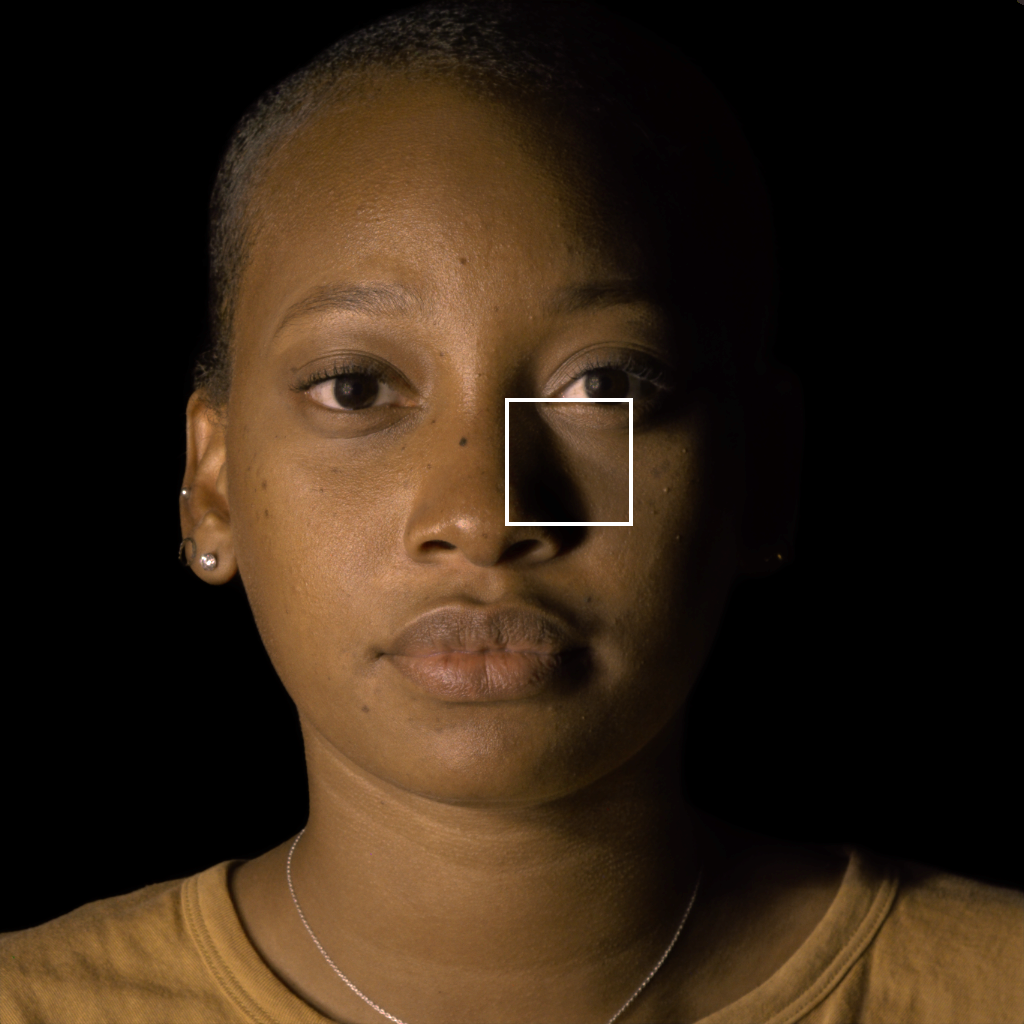}&
\includegraphics[width=\cropwidth]{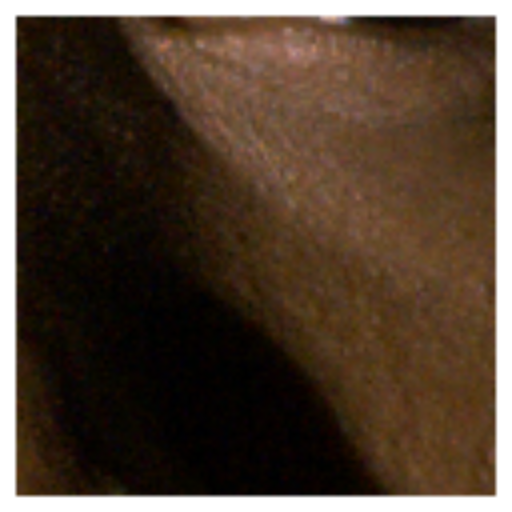}&
\includegraphics[width=\cropwidth]{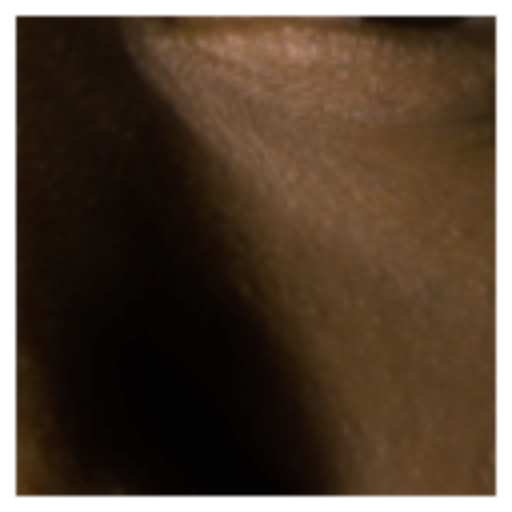}&
\includegraphics[width=\cropwidth]{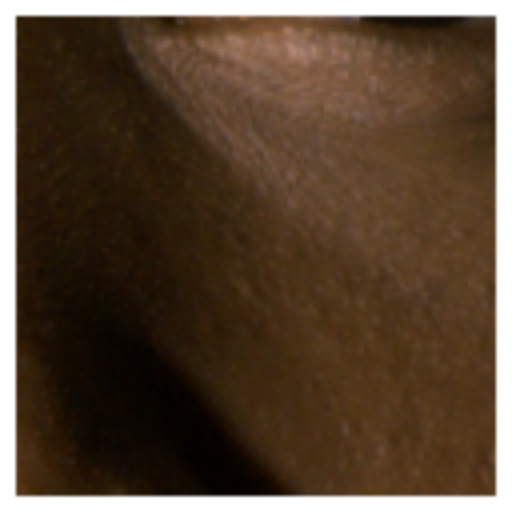}&
\includegraphics[width=\cropwidth]{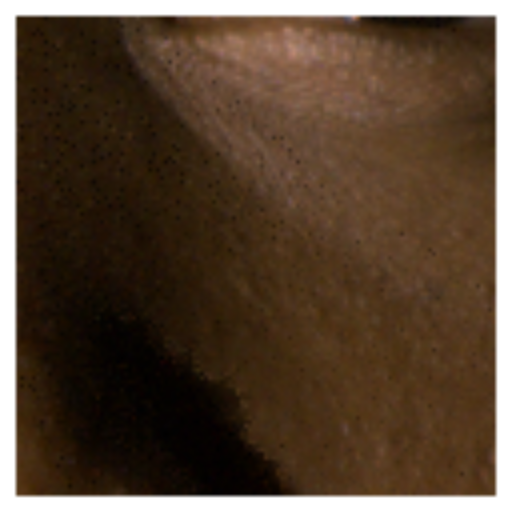}&
\includegraphics[width=\cropwidth]{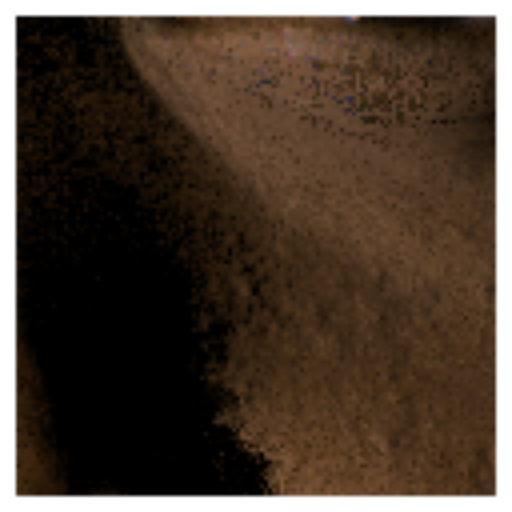}&
\includegraphics[width=\cropwidth]{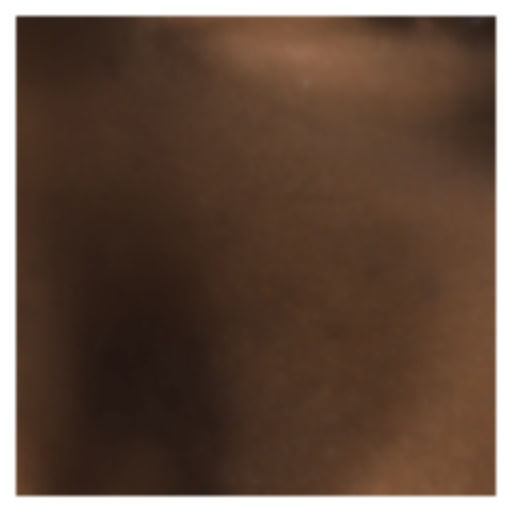}&
\includegraphics[width=\cropwidth]{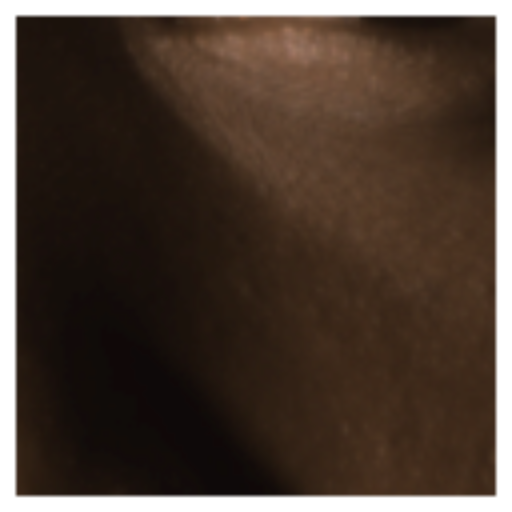}&
\includegraphics[width=\cropwidth]{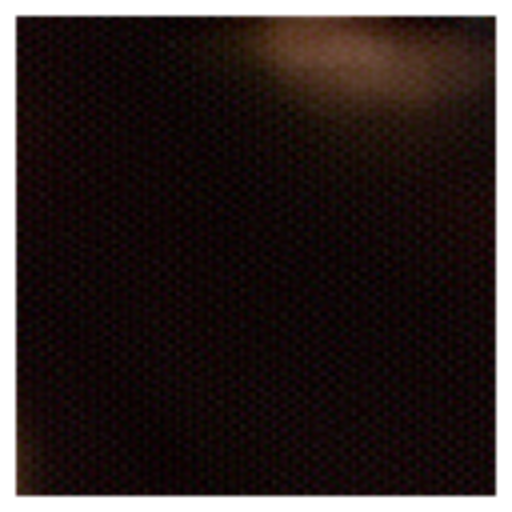}\\
\includegraphics[width=\cropwidth]{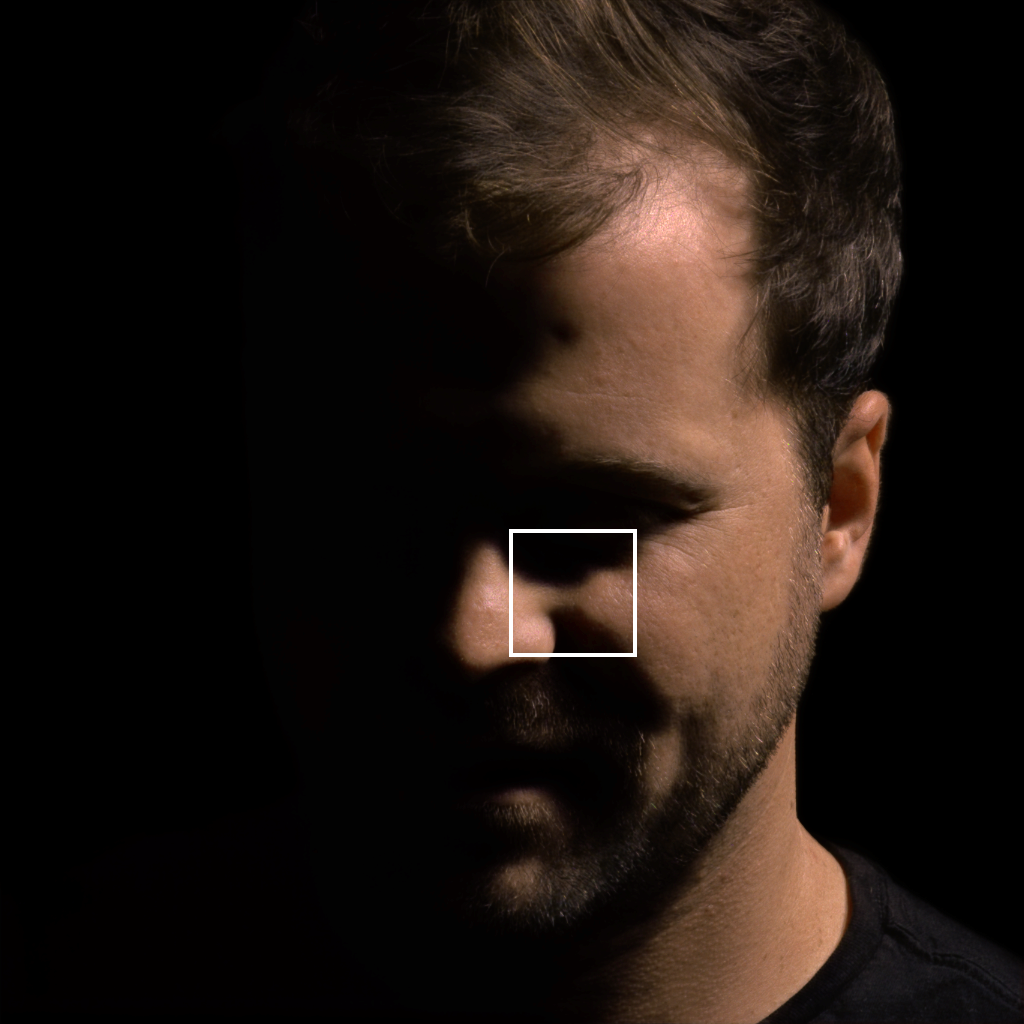}&
\includegraphics[width=\cropwidth]{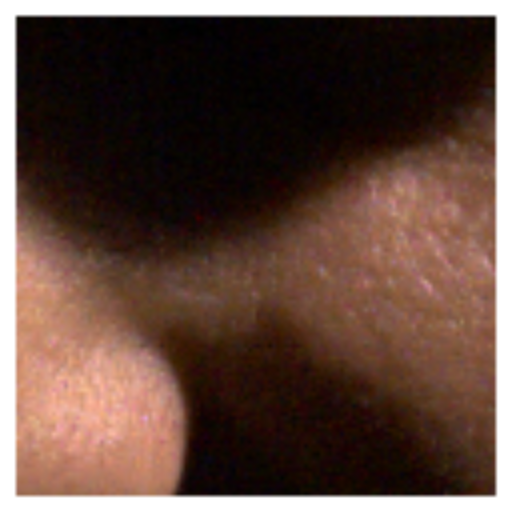}&
\includegraphics[width=\cropwidth]{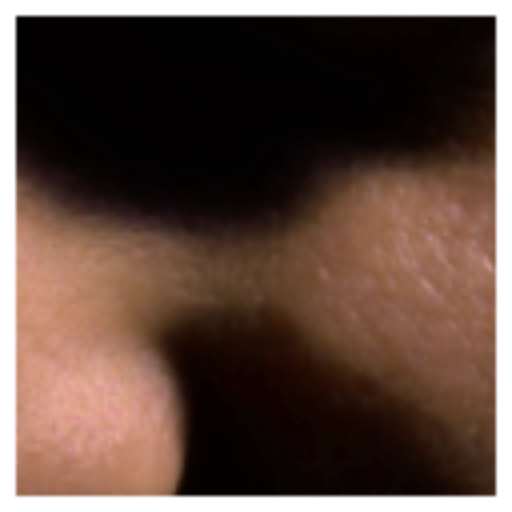}&
\includegraphics[width=\cropwidth]{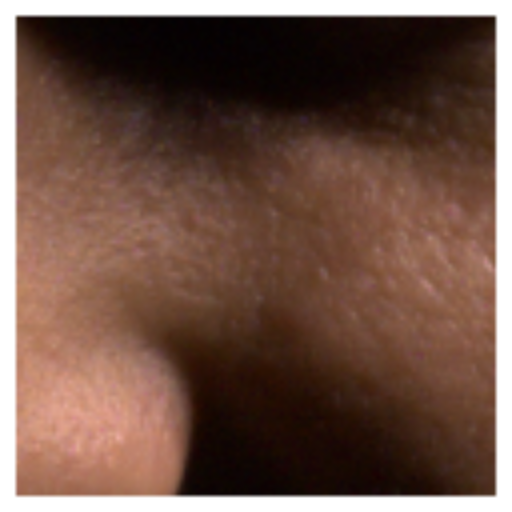}&
\includegraphics[width=\cropwidth]{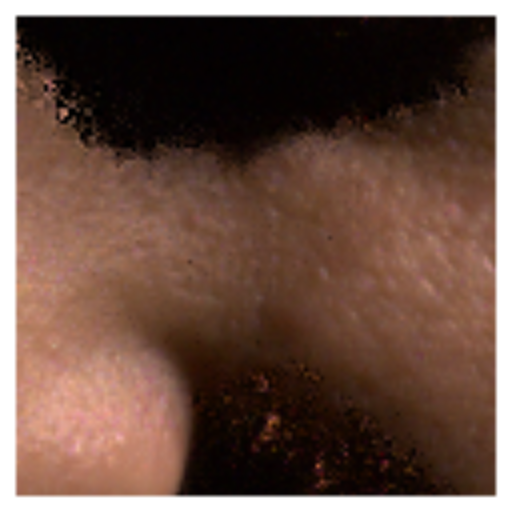}&
\includegraphics[width=\cropwidth]{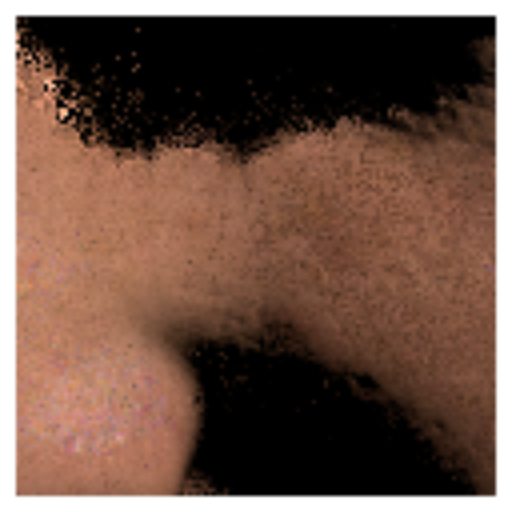}&
\includegraphics[width=\cropwidth]{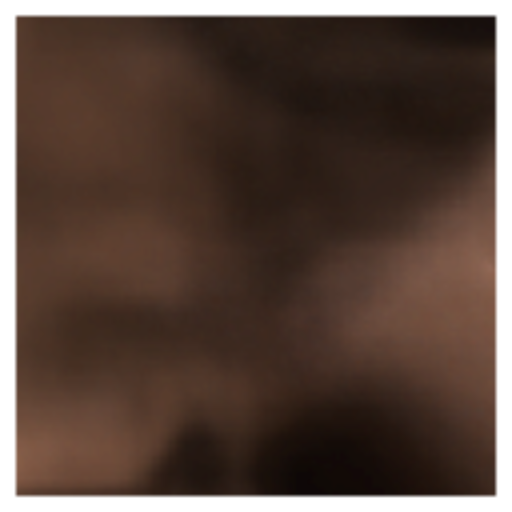}&
\includegraphics[width=\cropwidth]{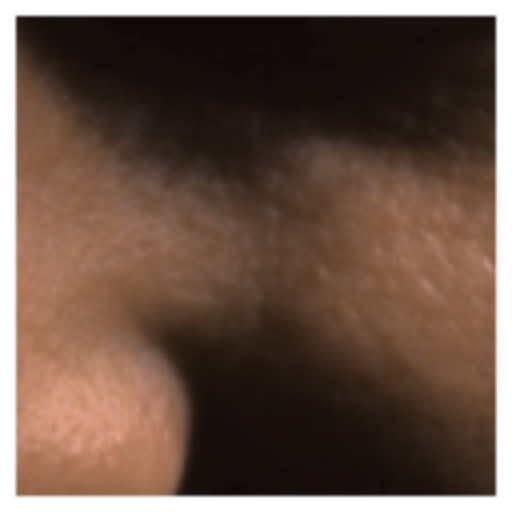}&
\includegraphics[width=\cropwidth]{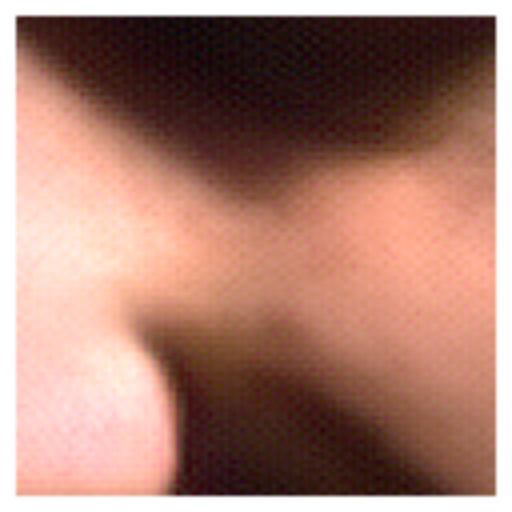}\\
\includegraphics[width=\cropwidth]{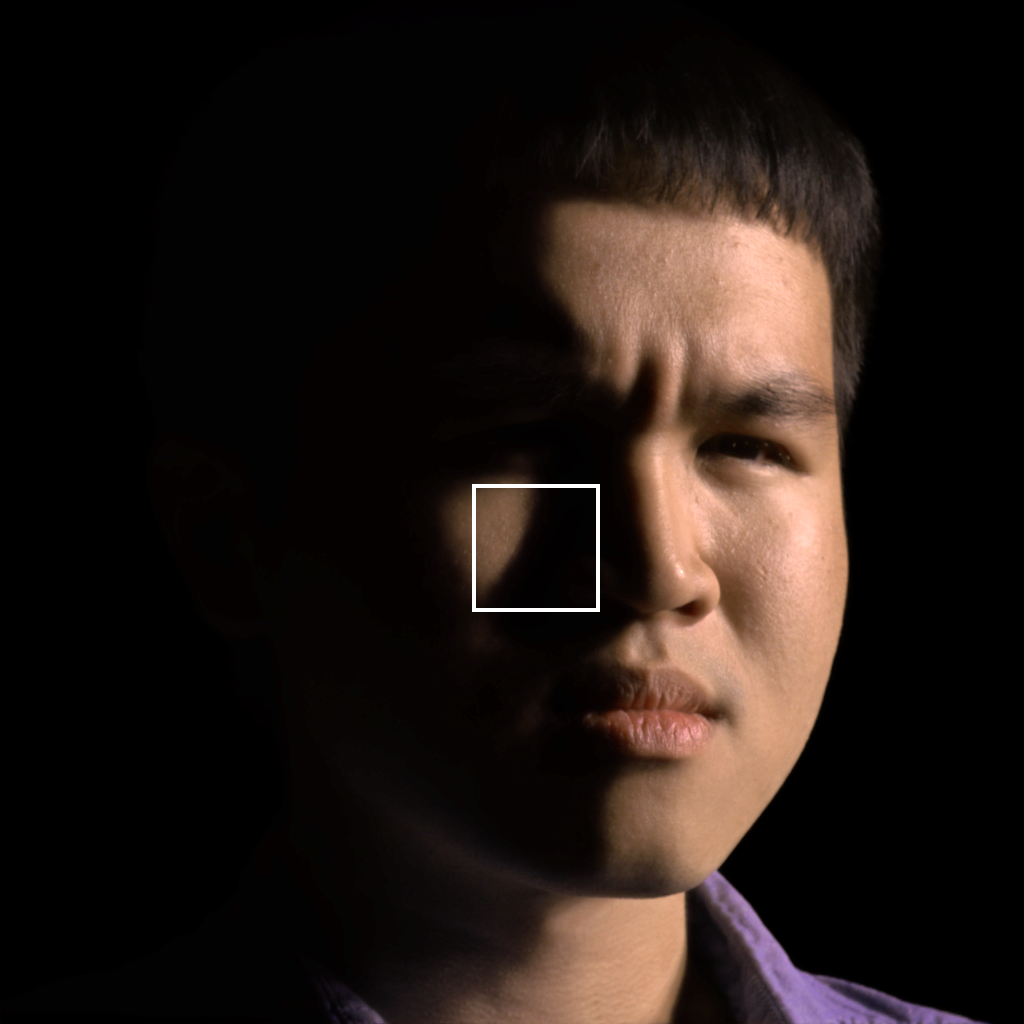}&
\includegraphics[width=\cropwidth]{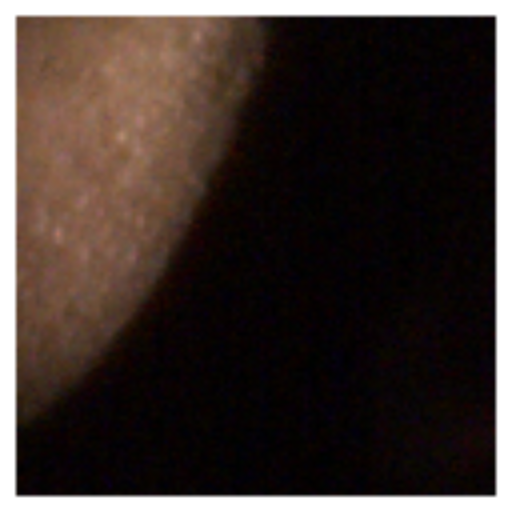}&
\includegraphics[width=\cropwidth]{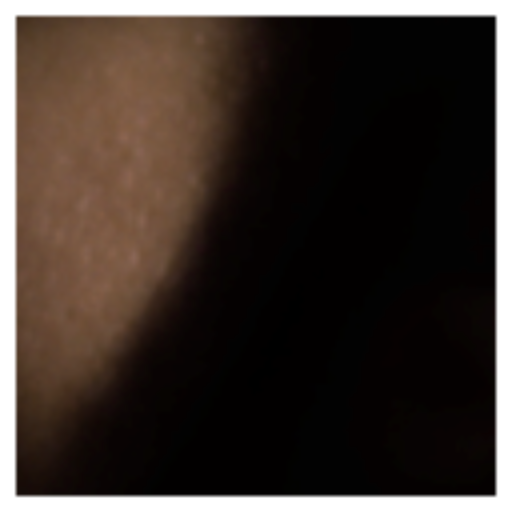}&
\includegraphics[width=\cropwidth]{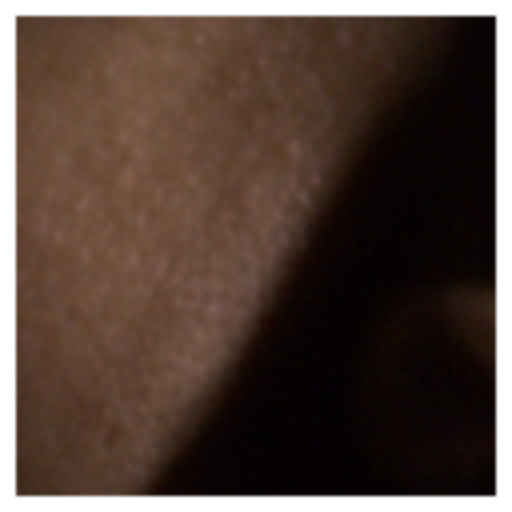}&
\includegraphics[width=\cropwidth]{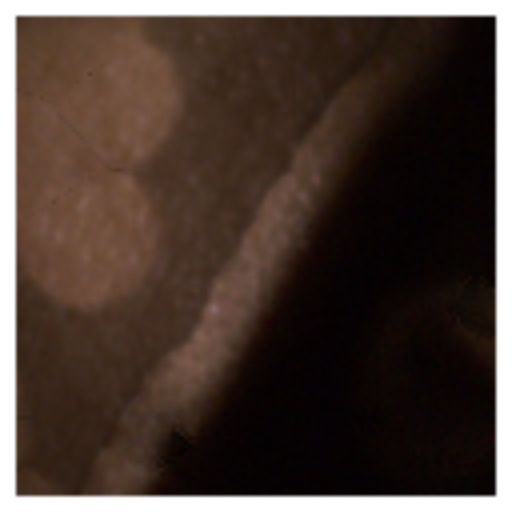}&
\includegraphics[width=\cropwidth]{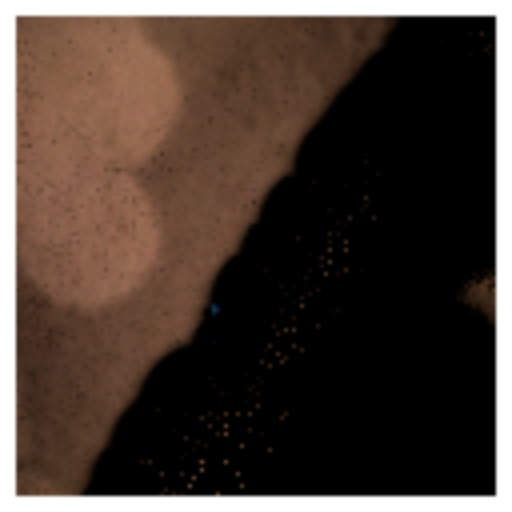}&
\includegraphics[width=\cropwidth]{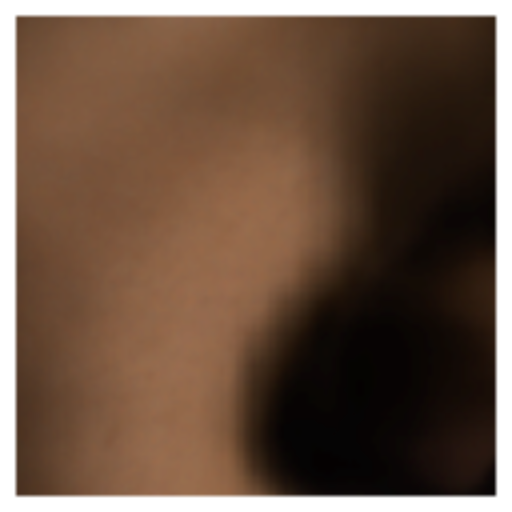}&
\includegraphics[width=\cropwidth]{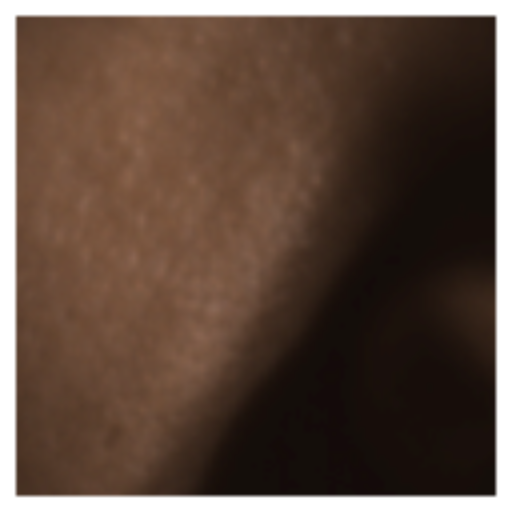}&
\includegraphics[width=\cropwidth]{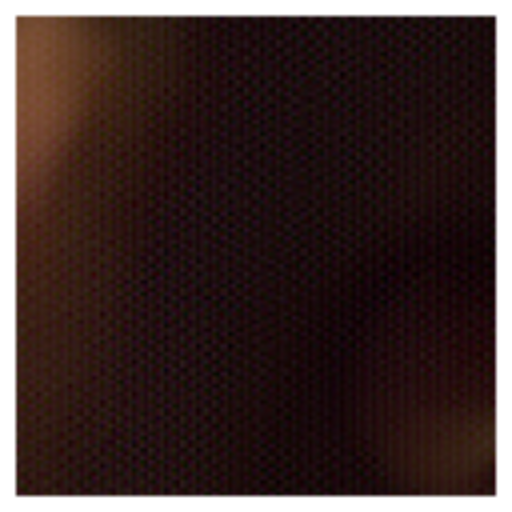}\\
\end{tabular}
% \vspace{-0.25cm}
\caption{
Here we present a qualitative comparison between our method and other light interpolation algorithms.
Traditional methods (linear blending, \citet{fuchs2007superresolution}, photometric stereo) retain detail but suffer from ghosting artifacts in shadowed regions.
Results from \citet{xu2018deep} and \citet{meka2019deep} exhibit significant oversmoothing and brightness changes.
Our method retains details and synthesizes shadows that resemble the ground truth.
}

\label{fig:related}
% \vspace*{-.1in}
\end{figure*}
\setlength{\tabcolsep}{6pt} % General space between cols (6pt standard)

\subsection{Ablation Study}
We first evaluate against ablated versions of our model, with results shown in Tab.~\ref{table:compare}.
In the ``Ours w/naive neighbors'' ablation we use the $k=8$ nearest neighbors in our active set during training. This setup leads to a match between our training and validation data, which results in better numerical performance (as shown in Tab.~\ref{table:compare}) but also significant overfitting: this apparent improvement in performance is misleading, because the validation set of our dataset has the same overly-regular sampling as the training set.
During our \emph{real} test-time scenario in which we synthesize with lights that do not lie on the regular hexagonal grid of our light stage, we see this ablated model generalizes poorly. In Fig.~\ref{fig:movingshadow} we visualize the output of our model and ablations of our model as a function of the query light direction. We see that our model is able to synthesize a cast shadow that is a smooth linear function in the image plane of the angle of the query light (after accounting for foreshortening, \etc). Ablations of our technique do not reproduce this linearly-varying shadow, due to the aliasing and overfitting problems described earlier. See the supplemental video for additional visualizations. 

In the ``Ours w/avg-pooling'' ablation we replace the alias-free pooling of our model with simple average pooling. As shown in Tab~\ref{table:compare}, ablating this component reduces performance. But more importantly, ablating this component also causes flickering during our \emph{real} test-time scenario in which we smoothly vary our light source, and this is not reflected in our quantitative evaluation. Because average pooling assigns a non-zero weight to images as they enter and exit our active set, renderings from this model will contain significant temporal instability. See the supplemental video for examples.
% In summary, alias-free weight ensures C0 continuity. And random active set selection push the model towards C1 continuity.
% \barron{I don't see why or how C1 continuity would be provided by dropout}

\subsection{Related Work Comparison}
We compare our results against related approaches that are capable of solving the relighting problem. The ``Linear blending'' baseline in Tab.~\ref{table:compare} produces competitive results, despite being a very simple algorithm: we simply blend the input images of our light stage according to our alias-free weights.
Because linear blending directly interpolates aligned pixel values, it is often able to retain accurate high frequency details in the flat region, and this strategy works well for minimizing our error metrics.
However, linear blending produces significant ghosting artifacts in shadows and highlights, as shown in Fig.~\ref{fig:related}. Though these errors are easy to detect visually, they appear to be hard to measure empirically.

We evaluate against the layer-based technique of \citet{fuchs2007superresolution} by decomposing an OLAT into diffuse, specular, and visibility layers, and interpolating the illumination individually for each layer. Although the method works well on specular objects as shown in the original paper, it performs less well on OLATs of human subjects, as shown in Tab.~\ref{table:compare}. This appears to be due to the complex specularities on human skin not being tracked accurately by the optical flow algorithm of \citet{fuchs2007superresolution}. Additionally, the interpolation of the visibility layer sometimes contains artifacts, which results in cast shadows being predicted incorrectly. That being said, the algorithm results in fewer ghosting artifacts than the linear blending algorithm, as shown in Fig.~\ref{fig:related} and as reflected by the E-LPIPS metric.

Using the layer decomposition produced by \citet{fuchs2007superresolution}, we additionally perform photometric stereo on the OLAT data by simple linear regression to estimate a per-pixel albedo image and normal map. Using this normal map and albedo image we can then use Lambertian reflectance to render a new diffuse image corresponding to the query light direction, which we add to the specular layer from~\cite{fuchs2007superresolution} to produce our final rendering.
As shown in Tab.~\ref{table:compare}, this approach underperforms that of \citet{fuchs2007superresolution}, likely due to the reflectance of human faces being non-Lambertian. Additionally, the scattering effect of human hair is poorly modeled in terms of a per-pixel albedo and normal vector. These limiting assumptions result in overly sharpened and incorrect shadow predictions, as shown in Fig.~\ref{fig:related}.
In contrast to this photometric stereo approach and the layer-based approach of \citet{fuchs2007superresolution}, our model does not attempt to factorize the human subject into a predefined reflectance model wherein interpolation can be explicitly performed. Our model is instead trained to identify a latent vector space of network activations in which naive linear interpolation results in accurate non-linearly interpolated images, which results in more accurate renderings.

The technique of \citet{xu2018deep} (retrained on our training data) represents another possible candidate for addressing our problem.
This technique does not natively solve our problem. In order to find the optimal lighting directions for relighting, it requires as input \emph{all} $302$ high-resolution images in each OLAT scan in the first step, which significantly exceeds the memory constraints of modern GPUs. To address this, we first jointly train the Sample-Net and the Relight-Net on our images (downsampled by a factor of 4$\times$ due to memory constraints) to identify 8 optimal directions from the 302 total directions of the light stage. 
Using those 8 optimal directions, we then retrain the Relight-Net using the full-resolution images from our training data, as prescribed in \citet{xu2018deep}. Table~\ref{table:compare} shows that this approach works poorly on our task. This may be because this technique is built around 8 fixed input images and is naturally disadvantaged compared to our approach, which is able to use any of the 302 light stage images as input. We therefore also evaluate a variant of \citet{xu2018deep} where we use the same active-set selection approach used by our model to select the images used to train their Relight-Net. By using our active-set selection approach (Sec.~\ref{sec:activeset}) this baseline is able to better reason about local information, which improves performance as shown in Tab.~\ref{table:compare}. However, this baseline still results in flickering artifacts when rendering with moving lights, because (unlike our approach) it is sensitive to the aliasing induced when images leave and enter the active set.
% Second, their method is designed for relighting from a hemisphere rather than a full sphere. Hence, our data is inherently more challenging. %This creates more difficulties for them when dealing with our data captured under a full spherical dome.

\setlength{\tabcolsep}{1pt} % General space between cols (6pt standard)
\renewcommand{\fullwidth}{0.175\textwidth} 
\renewcommand{\cropwidth}{0.13\textwidth} 
\begin{figure*}[!ht]
\centering
\begin{tabular}[t]{ccccccc}
$n = 100$ & $n = 150$ & $n = 200$ & $n = 250$ & $n = 302$ & Groundtruth & Groundtruth (Complete) \\
\includegraphics[width=\cropwidth]{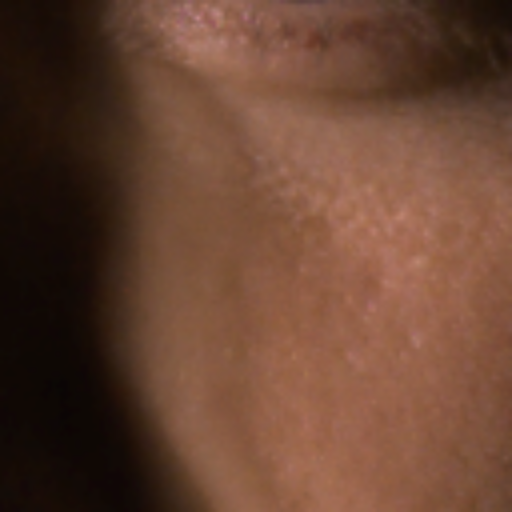}&
\includegraphics[width=\cropwidth]{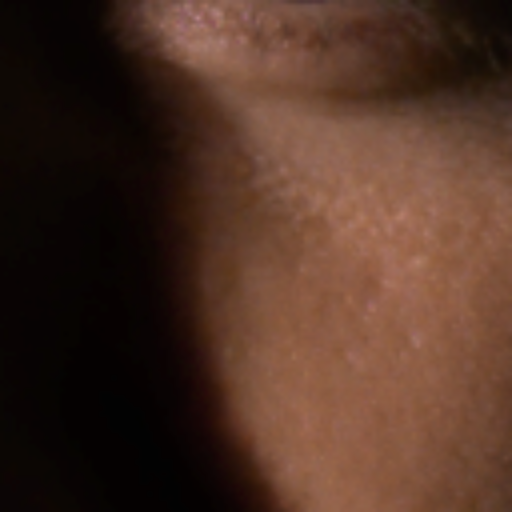}&
\includegraphics[width=\cropwidth]{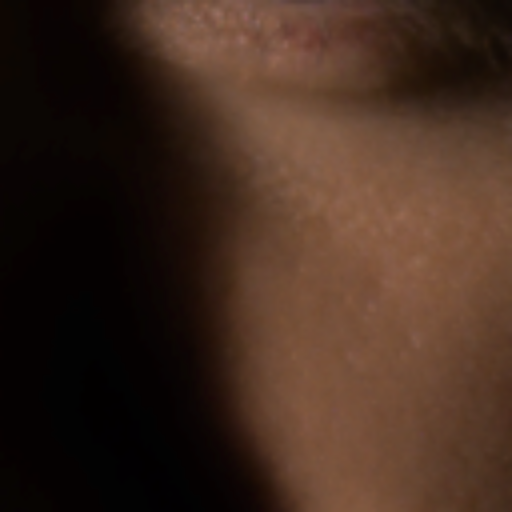}&
\includegraphics[width=\cropwidth]{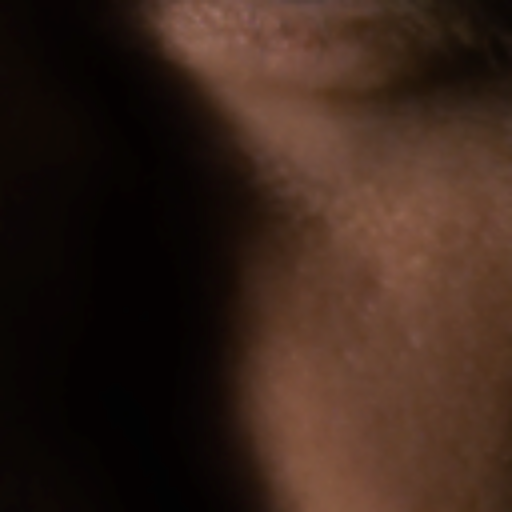}&
\includegraphics[width=\cropwidth]{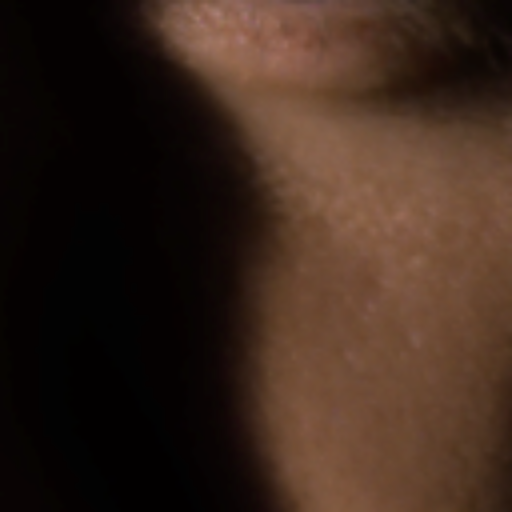}&
\includegraphics[width=\cropwidth]{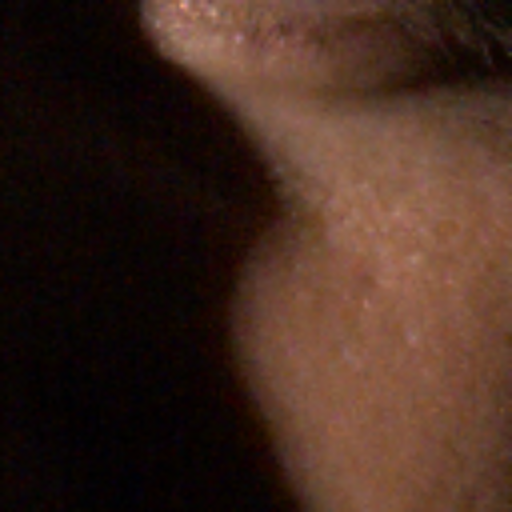} & 
\multirow{4}{*}[6.5em]{\includegraphics[trim=100 0 100 0, clip, width=\fullwidth]{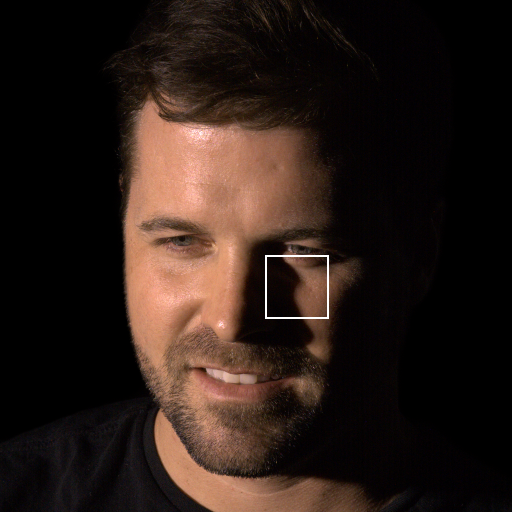}}\\
\multicolumn{6}{c}{\small (a) Ours} & \\
\includegraphics[width=\cropwidth]{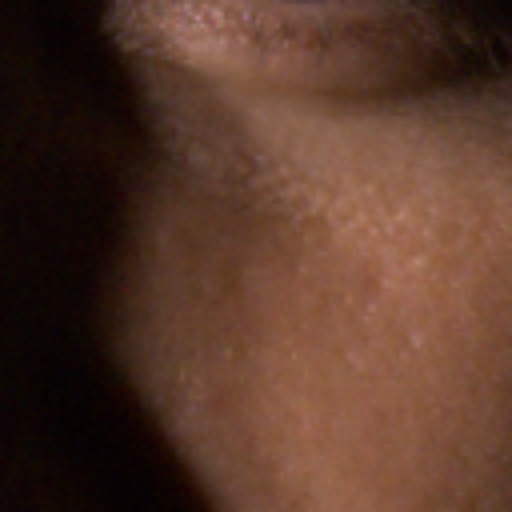}&
\includegraphics[width=\cropwidth]{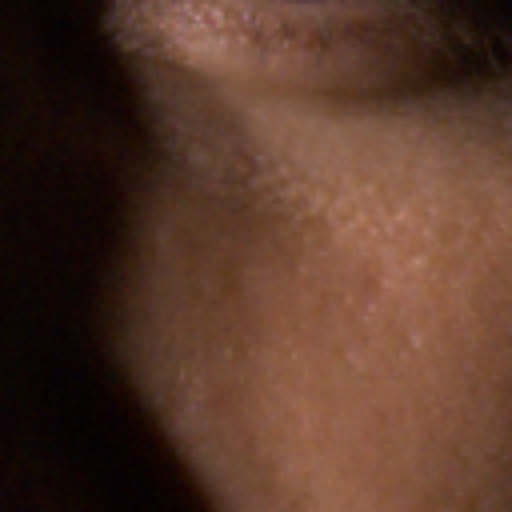}&
\includegraphics[width=\cropwidth]{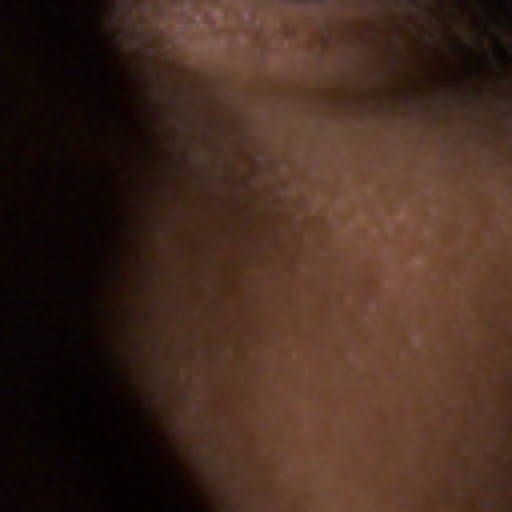}&
\includegraphics[width=\cropwidth]{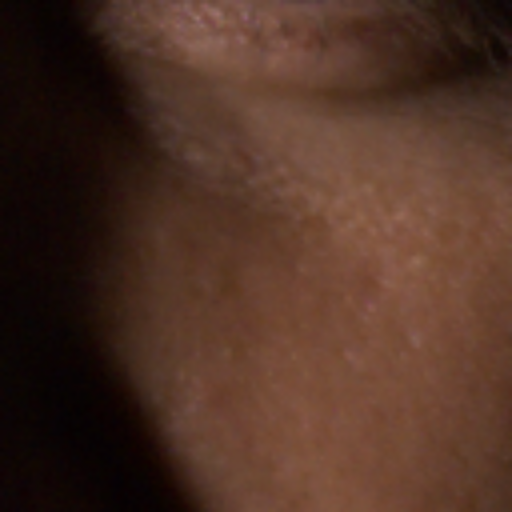}&
\includegraphics[width=\cropwidth]{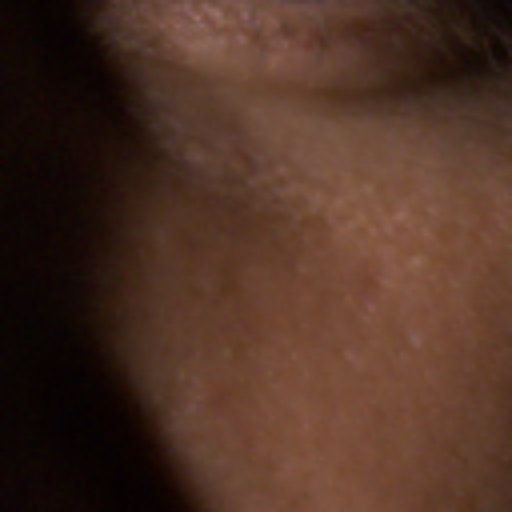}&
\includegraphics[width=\cropwidth]{image/sparse/crop_0_gt.png} & \\
\multicolumn{6}{c}{\small (b) Linear Blending} & \\
\end{tabular}
% \vspace{-0.25cm}
\caption{Here we compare the performance of our model against linear blending as we reduce $n$, the number of lights in our light stage.
As we decrease the number of available lights from $n=302$ to $n=100$, the quality of our model's rendered shadow degrades slowly. Linear blending, in contrast, is unable to produce an accurate rendering even with access to all lights.
}
\label{fig:stagesubsample}
\end{figure*}

\setlength{\tabcolsep}{6pt} % General space between cols (6pt standard)

\begin{figure}
    \centering
    \includegraphics[width=\linewidth]{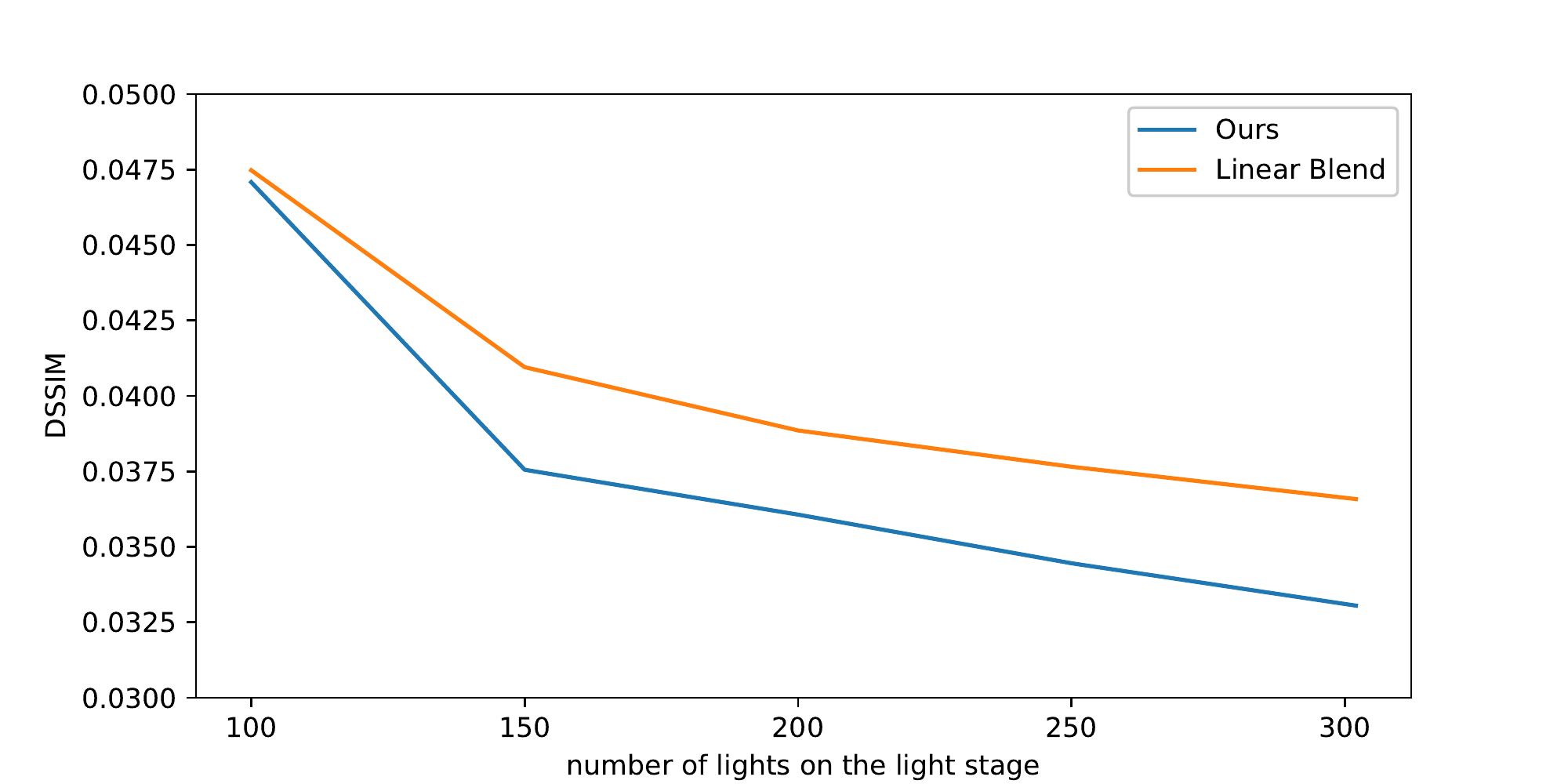}
    % \vspace*{-.15in}
    \caption{
    The image quality of relighting algorithms will gradually reduce as we remove lights from the light stage. However, our algorithm is able to retain the image quality to a greater extent with fewer lights compared to naive linear blending.
    }
    \label{fig:sparse_plot}
    % \vspace*{-.2in}
\end{figure}

\setlength{\tabcolsep}{1pt} % General space between cols (6pt standard)
\renewcommand{\fullwidth}{0.219\textwidth} 
\renewcommand{\cropwidth}{0.27\textwidth} 
\begin{figure*}[!ht]
\centering
\begin{tabular}[t]{cccc}
\makecell{Captured image under light A} & \multicolumn{2}{c}{$\xleftarrow{\hspace*{1cm}}$ Interpolation between captured lights $\xrightarrow{\hspace*{1cm}}$} & \makecell{Captured image under light B} \\
\multirow{8}{*}[2.9em]{\includegraphics[width=\fullwidth]{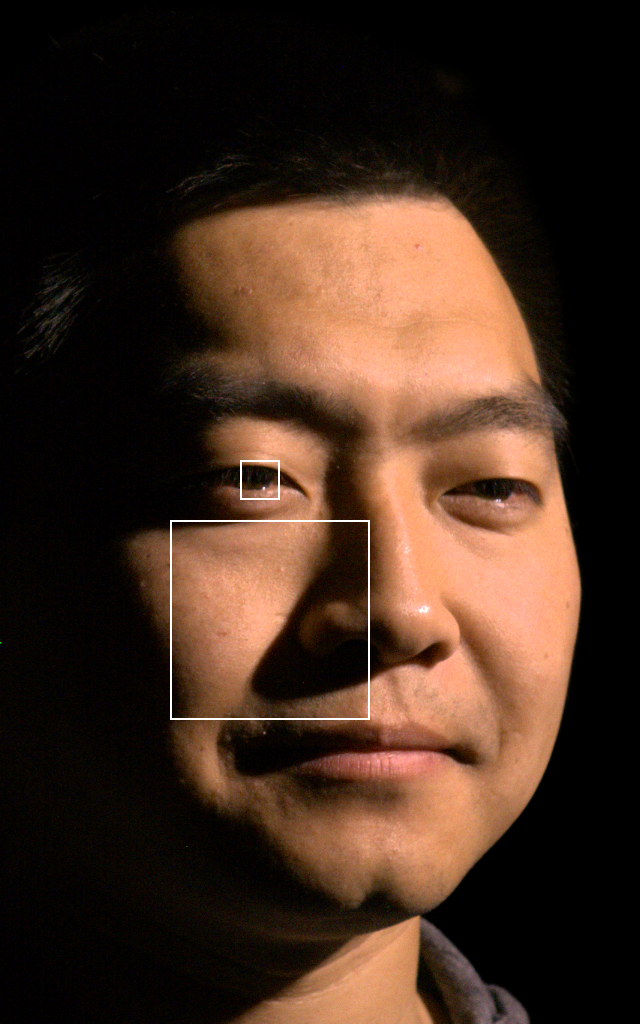}}&
\includegraphics[width=\cropwidth]{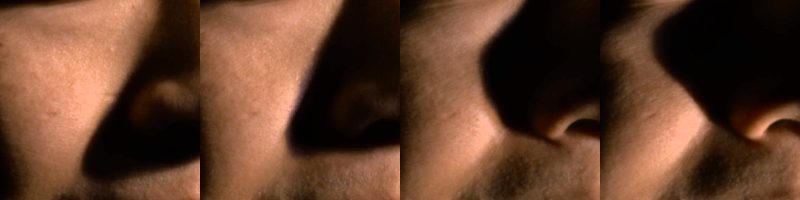}&
\includegraphics[width=\cropwidth]{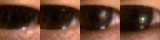}&
\multirow{8}{*}[2.9em]{\includegraphics[width=\fullwidth]{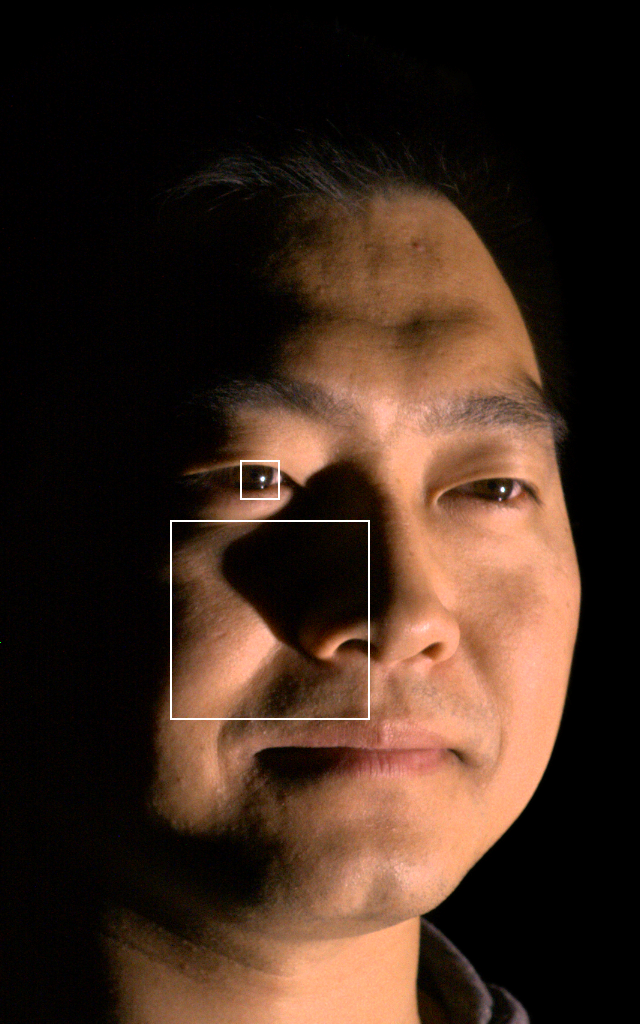}}\\
& \multicolumn{2}{c}{\small (a) Ours}\\
& 
\includegraphics[width=\cropwidth]{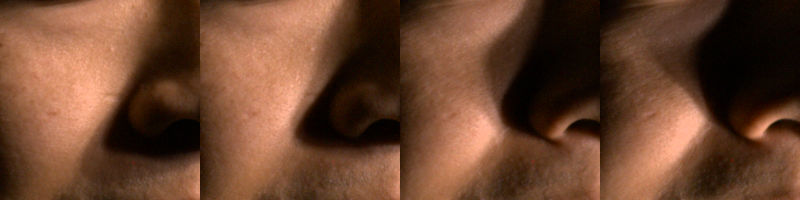}&
\includegraphics[width=\cropwidth]{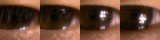}&\\
& \multicolumn{2}{c}{\small (b) Linear Blending}\\
& 
\includegraphics[width=\cropwidth]{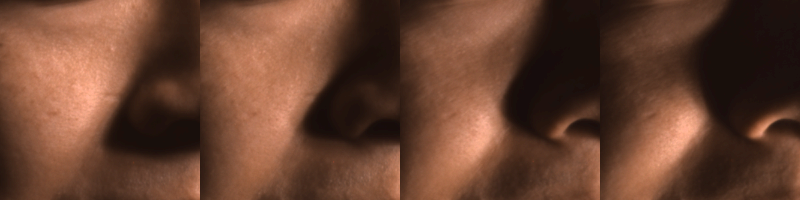}&
\includegraphics[width=\cropwidth]{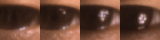}&\\
& \multicolumn{2}{c}{\small (c) \cite{xu2018deep} w/ adaptive sampling}\\
& 
\includegraphics[width=\cropwidth]{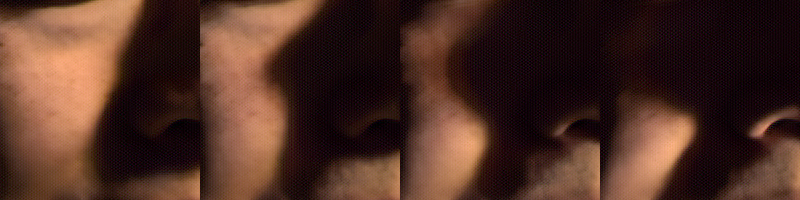}&
\includegraphics[width=\cropwidth]{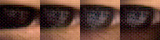}&\\
& \multicolumn{2}{c}{\small (d) \cite{meka2019deep}}\\
\end{tabular}
% \vspace{-0.25cm}
\caption{
Here we use produce interpolated images corresponding to ``virtual'' lights between two real lights of the light stage.
Our model (a) produces renderings where sharp shadows and accurate highlights move realistically. Linear blending (b) and \citet{xu2018deep} with adaptive sampling result in ghosting artifacts and duplicated highlights. The results from \citet{meka2019deep} contain blurry highlights and shadows with unrealistic motion.
}
% \vspace{-0.3cm}
\label{fig:interpolation}
\end{figure*}

We also evaluate Deep Reflectance Fields~\cite{meka2019deep} for our task, which is also outperformed by our model. This is likely because their model is specifically designed for fast and approximate video relighting and uses only two images as input, while our model has access to the entire OLAT scan and is designed to prioritize high-quality rendering.

\subsection{Light Stage Subsampling}
An interesting question in light transport acquisition is how many images (light samples) are needed to reconstruct the full light transport function.
To address this question, we present an experiment in which we remove some lights from our training set and use only this subsampled data during training and inference. We reduce the number of lights on the light stage $n$ (while maintaining a uniform distribution on the sphere) to $[250, 200, 150, 100]$, while also changing the number of candidates $m$ and the active set size $k$ to $[14, 12, 10, 8]$ and $[7, 6, 5, 4]$ respectively.
%We remove the light that is closest to its neighboring light one by one, in order to keep the light distribution uniform on the sphere.
Image quality on the complete validation dataset (with all 302 lights) as a function of the number of subsampled training/input lights is shown in Fig.~\ref{fig:sparse_plot}.
As expected, relighting quality decreases as we remove the lights, but we see that the rendering quality of our method decreases more slowly than that of linear blending.
This can also be observed in Fig.~\ref{fig:stagesubsample}, where we present relit renderings using these subsampled light stages.
We see that removing lights reduces accuracy compared to the ground truth, but that our synthesized shadows remain relatively sharp:
ghosting artifacts only appear when $n=100$.
In comparison, linear blending produces ghosting artifacts near shadow boundaries for all values of $n$.
During test time, our model can also produce accurate shadows and sharp highlights. Please refer to our supplementary video for our qualitative comparison.

%% file: 5-app.tex
\section{Continuous High-Frequency Relighting}\label{sec:app}
% \ravi{The introduction made a big deal of continuous relighting by creating 
% a light stage with thousands of virtual lights, from which you could do 
% high-frequency environment relighting.  Maybe you're intending that in this 
% section, but you should explicitly return to it, and maybe the intro can even 
% ref the corresponding figures in this section.  You should also refer to your
% video.}
A key benefit of our method is the ability to "super-resolve" an OLAT scan with virtual lights at a higher resolution than the original light stage data, thereby enabling continuous high-frequency relighting with an essentially continuous lighting distribution (or equivalently, with a light stage whose sampling frequency is unbounded).  In this section, we present three applications of this idea.

\paragraph{Precise Directional Light Relighting}
Traditional image-based relighting methods produce accurate results near the observed lights of the stage, but may introduce ghosting effects or inaccurate shadows when no observed light is nearby.
In Fig.~\ref{fig:interpolation} we try to interpolate the image between two lights on the stage. As shown in the second and the third row, linear blending or \citet{xu2018deep} with adaptive sampling does not produce realistic results and always contains multiple superposed shadows or highlights. The shadows produced by \citet{meka2019deep} are sharp, but are not moving smoothly when the light moves. In contrast, our method is able to produce sharp and realistic images for arbitrary light directions: highlights and cast shadows move smoothly as we change the light direction, and our results have comparable sharpness to the (non-interpolated) groundtruth images that are available.
%In other words, our method is able to produce sharp and realistic images lit by arbitrary light directions.

\begin{figure}[!t]
    \centering
    \begin{subfigure}[b]{.49\linewidth}
    \centering
    \includegraphics[width=\linewidth]{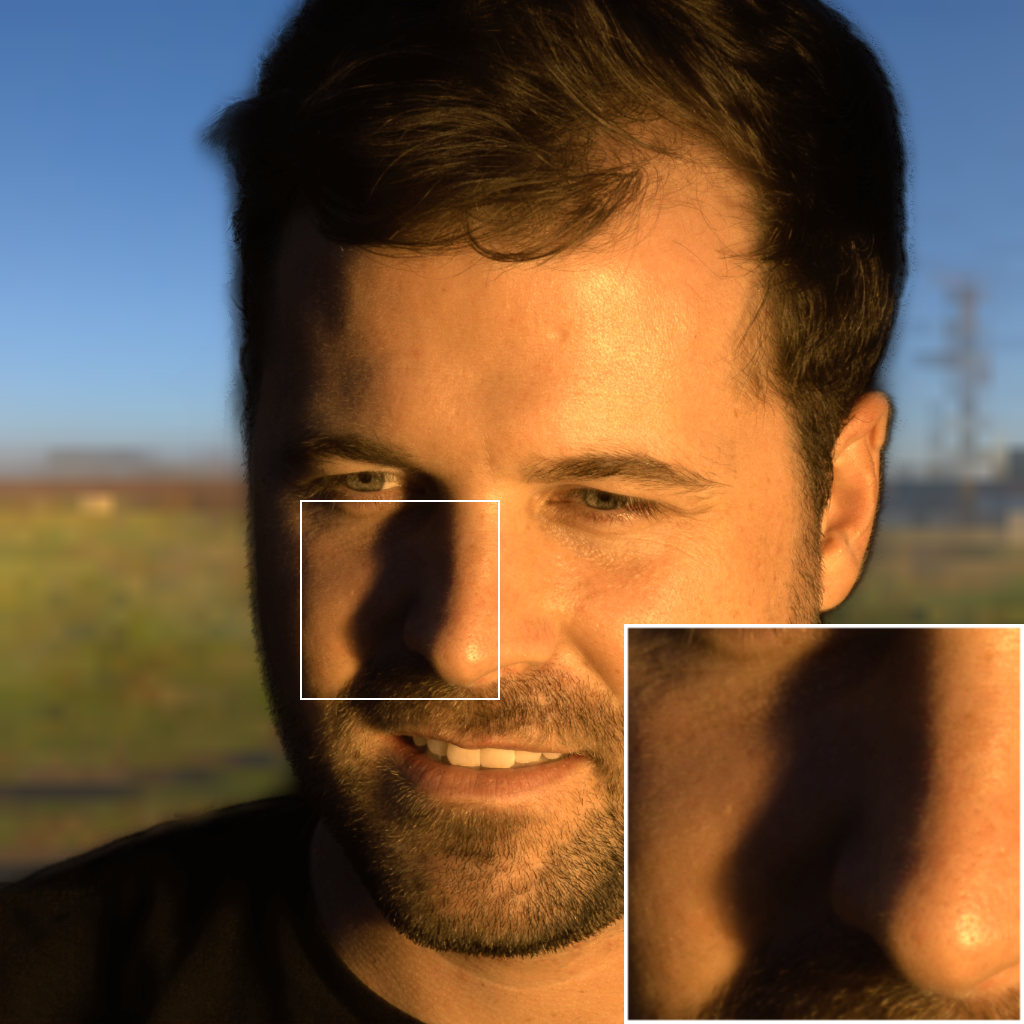}
    \caption{With super-resolution}
    \end{subfigure}
    \begin{subfigure}[b]{.49\linewidth}
    \centering
    \includegraphics[width=\linewidth]{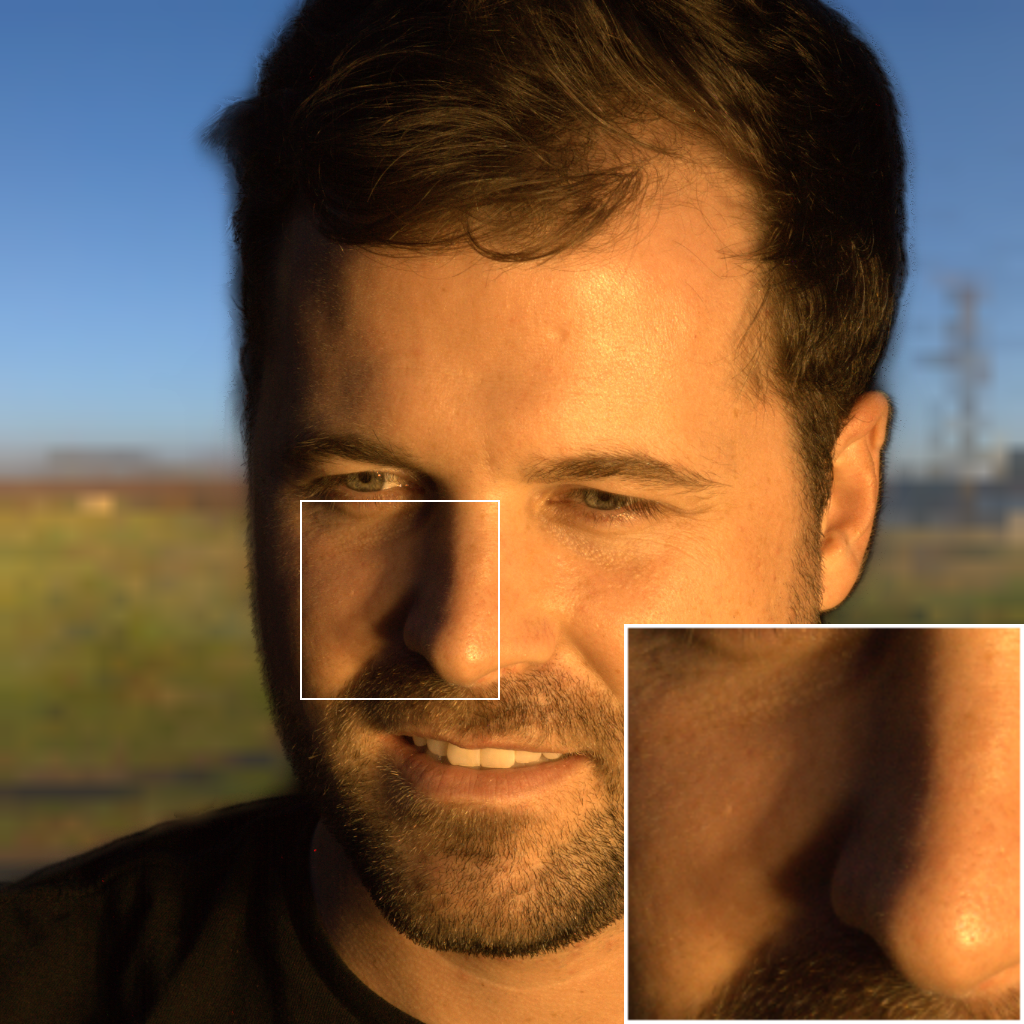}
    \caption{Without super-resolution}
    \end{subfigure}
    % \includegraphics[width=0.49\linewidth]{image/envmap/ours.png}
    % \includegraphics[width=0.49\linewidth]{image/envmap/linear.png}
    % \vspace{-0.25cm}
    \caption{
    Our model (a) is able to produce accurate relighting results under high-frequency environments by super-resolving the light stage before performing image-based relighting~\cite{debevec2000acquiring}.
    Using the light stage data as-provided (b) results in ghosting.
    }
    \label{fig:envmap}
    % \vspace*{-.2in}
\end{figure}

\begin{figure}[!t]
    \centering
    \includegraphics[width=0.99\linewidth]{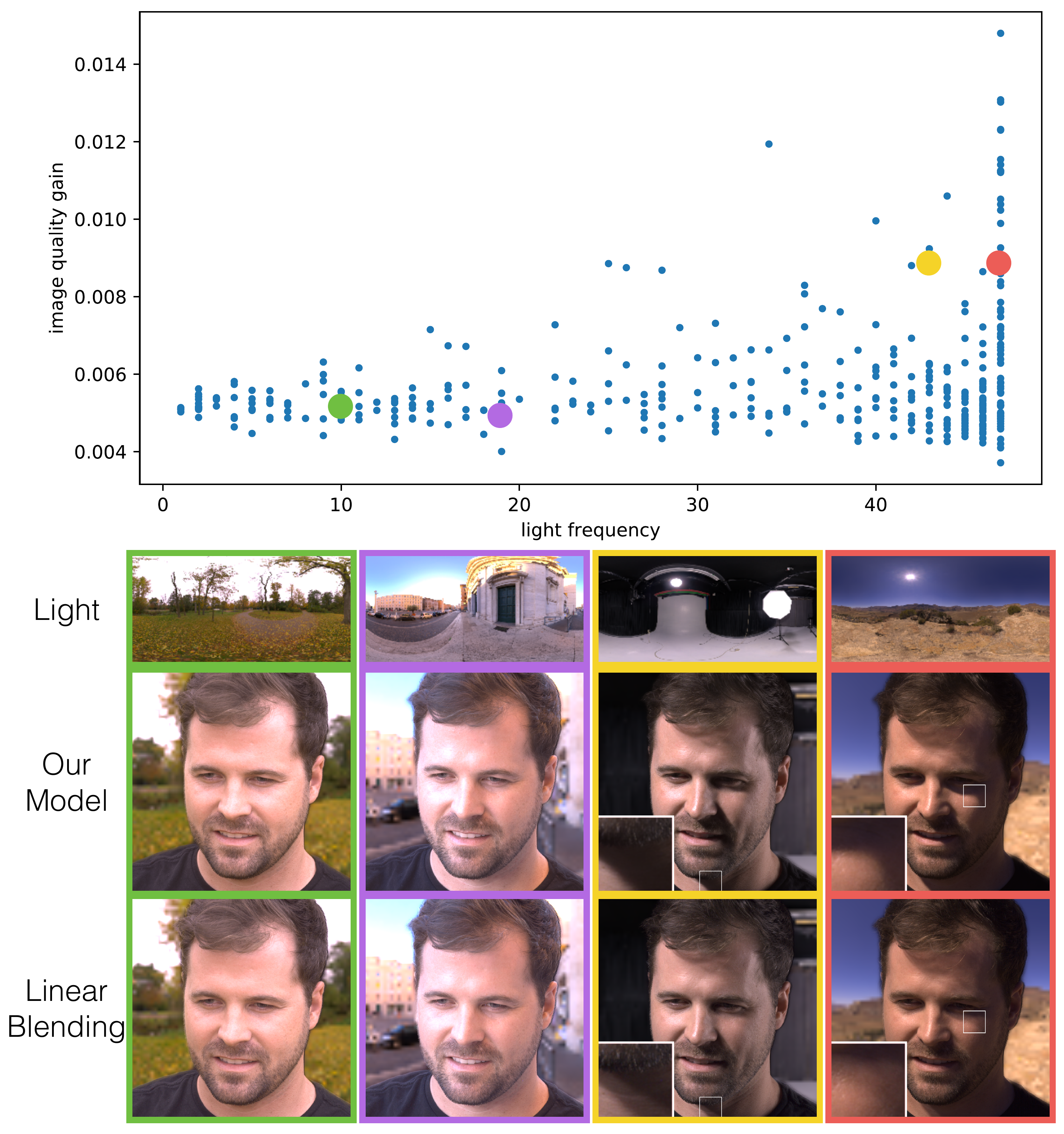}
    % \vspace{-0.25cm}
    \caption{
    In the top figure, each blue dot represents a lighting environment. We render a portrait under this environment using both linear-blending and our method, and measure the image difference using SSIM to evaluate the quality gain of our algorithm. The image quality improvement produced by our model becomes more apparent when the environment map has more high-frequency variation. In the bottom figure, we compare the rendered images using our model and linear blending under environment maps with different frequencies. Our model produces similar results to linear blending when the lighting variation is low frequency (left columns). As the lighting variation becomes higher frequency, our model produces better renderings with fewer artifacts and sharper shadows (right columns).
    }
    \label{fig:freq}
    % \vspace*{-.2in}
\end{figure}

\paragraph{High Frequency Environment Relighting}
OLAT scans captured from a light stage can be linearly blended to reproduce images that appear to have been captured under a specific environment. The pixel values of the environment map are usually distributed to the nearest or neighboring lights on the light stage for blending. This traditional approach may cause ghosting artifacts in shadows and specularities, due to the finite sampling of light directions on the light stage. Although this ghosting is hardly noticeable when the lighting is low-frequency, it can be significant when the environment contains high frequency lighting, such as the sun in the sky.
These ghosting artifacts can be ameliorated by using our model. Given an environment map, our algorithm can predict the image corresponding to the light direction of each pixel in the environment map. By taking a linear combination of all such images (weighted by their pixel values and solid angles), we are able to produce a rendering that matches the sampling resolution of the environment map.
% We can thus suppress the ghosting artifacts to be minimal.
As shown in Fig.~\ref{fig:envmap}, this approach produces images with sharp shadows and minimal ghosting when given a high-frequency environment, while linear blending does not. In this example, we use an environment resolution of $256 \times 128$, which corresponds to a super-resolved light stage with \num[group-separator={,}]{32768} lights.  Please see our video for more environment relighting results.

We now analyze the relationship between the image quality gain from our model and the frequency of the lighting. Specifically, we evaluate for which environments, and at what frequencies, our algorithm will be required for accurate rendering, and conversely how our model performs in low-frequency lighting environments where previous solutions are adequate.  For this purpose, we use one OLAT scan, and render it under 380 high quality indoor and outdoor environment maps (environments downloaded from hdrihaven.com) using both our model and linear blending.  We then measure the image quality gain from our model by computing the DSSIM value between our rendering and that from linear blending.  We measure the frequency of the environmental lighting by decomposing it into spherical harmonics (up to degree 50), and finding the degree below which 90\% of the energy can be recovered.  

As shown in Fig.~\ref{fig:freq}, the benefit of using our model becomes larger when the frequency of the environment increases. For low-frequency light (up to degree 15 spherical harmonics), our model produces almost identical results compared to the traditional linear blending method. This is a desired property, showing that our method reduces gracefully to linear blending for low frequency lighting, and thus produces high quality results for any low or high-frequency environment.  
As the frequency of the lighting becomes higher, the renderings of our model contain sharper and more accurate shadows without ghosting artifacts.  Note that there is some variation among the environment maps as expected; even a very high-frequency environment could coincidentally have its brightest lights aligned with one of the light in the light stage, leading to low error in linear blending and comparable results to our method.  Nevertheless, the trend is clear in Fig.~\ref{fig:freq} with many high-frequency environments requiring our algorithm for lighting super-resolution.  

According to the plot, we conclude that our model is necessary when the light frequency is equal or larger than about 20, which means more than $21^2 = 441$ basis functions are needed to recover the lighting. This number has the same order as the number of lights in the stage ($n=302$). This observation agrees with intuition and frequency analysis. If the environment cannot be recovered using the limited lighting basis in the light stage, then light super-resolution is needed to generate new bases in order to accurately render the shadow and highlights.

\setlength{\tabcolsep}{1pt} % General space between cols (6pt standard)
\renewcommand{\fullwidth}{0.136\textwidth} 
\renewcommand{\cropwidth}{0.107\textwidth} 
\begin{figure}[!t]
\centering
\begin{tabular}[t]{@{}cccc@{}}
Our full image & \multicolumn{3}{c}{ Increased shadow radius $\xrightarrow{\hspace*{1cm}}$ } \\
\multirow{4}{*}[5.25em]{\includegraphics[trim=50 0 400 0, clip, width=\fullwidth]{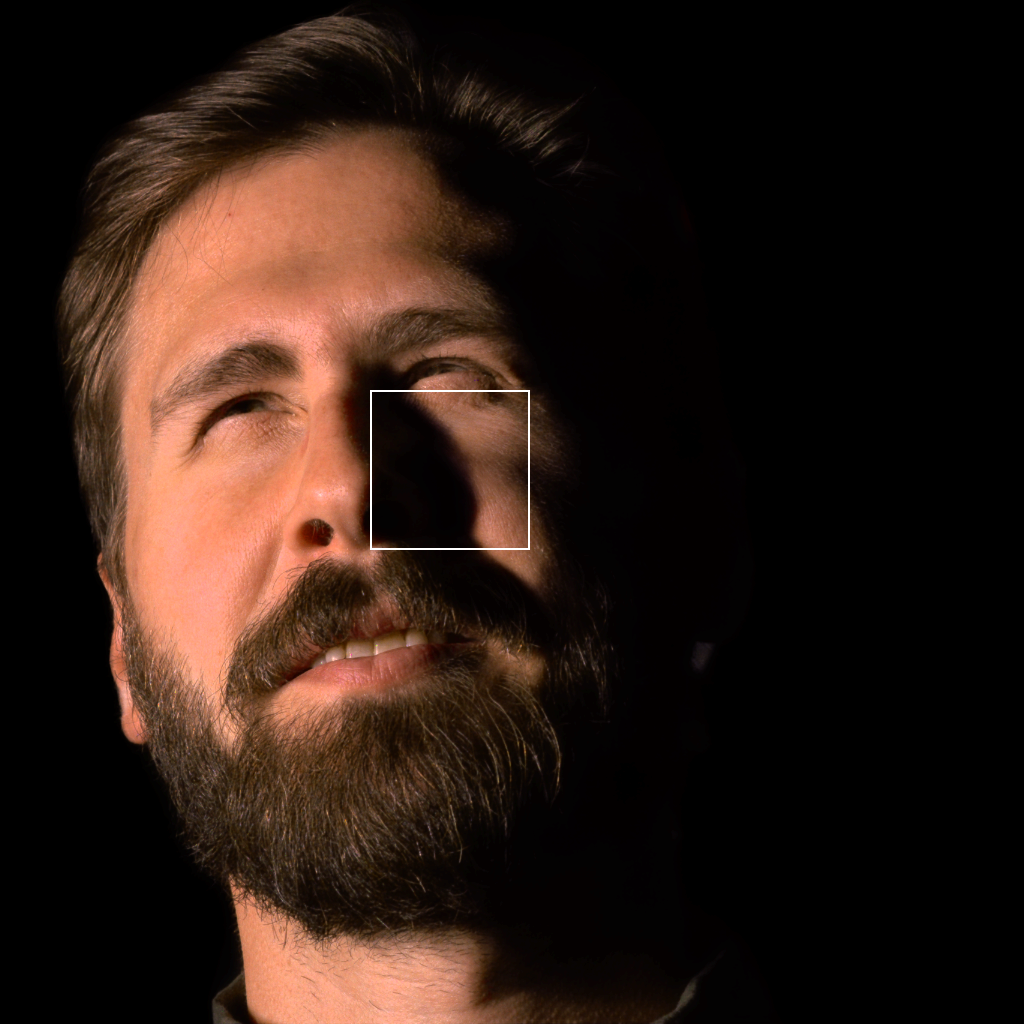}}&
\includegraphics[width=\cropwidth]{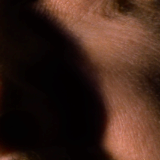}&
\includegraphics[width=\cropwidth]{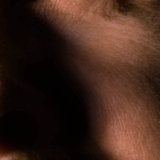}&
\includegraphics[width=\cropwidth]{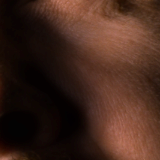}\\
& \multicolumn{3}{c}{\small (a) Our Model}\\
& \includegraphics[width=\cropwidth]{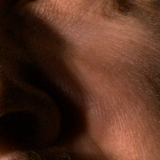}&
\includegraphics[width=\cropwidth]{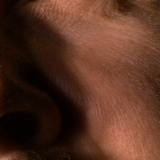}&
\includegraphics[width=\cropwidth]{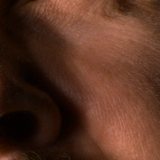}\\
& \multicolumn{3}{c}{\small (b) Linear Blending}\\
\end{tabular}
% \vspace{-0.25cm}
\caption{
Soft shadows can be rendered by synthesizing and averaging images corresponding to directional light sources within some area on the sphere.
Soft shadows rendered by our method (a) are more realistic and contain fewer ghosting artifacts than those rendered using linear blending (b).
}
\label{fig:softness}
% \vspace*{-.1in}
\end{figure}

\paragraph{Lighting Softness Control}
% \changed{
Our model's ability to render images under arbitrary light directions also allows us to control the softness of the shadow.
Given a light direction, we can densely synthesize images corresponding to the light directions around it, and average those images together to produce a rendering with realistic soft shadows (the sampling radius of these lights determines the softness of the resulting shadow). As shown in Fig.~\ref{fig:softness}, our model is able to synthesize realistic shadows with controllable softness, which is not possible using traditional linear blending methods.

%% file: 6-conclude.tex
\section{Conclusions and Future Work}
\label{sec:conclude}

The light stage is a crucial tool for enabling the image-based relighting of human subjects in novel environments.
But as we have demonstrated, light stage scans are undersampled with respect to the angle of incident light, which means that synthesizing virtual lights by simply combining images results in ghosting on shadows and specular highlights.
We have presented a learning-based solution for super-resolving light stage scans, thereby allowing us to create a ``virtual'' light stage with a much higher angular lighting resolution, which allows us to render accurate shadows and highlights in high-frequency environment maps.
Our network works by embedding input images from the light stage into a learned space where network activations can then be averaged, and decoding those activations according to some query light direction to reconstruct an image.
In constructing this model, we have identified two critical issues: an overly regular sampling pattern in light stage training data, and aliasing introduced when pooling activations of a set of nearest neighbors. These issues are addressed through our use of a dropout-like supersampling of neighbors in our active set, and our alias-free pooling technique.
By combining ideas from conventional linear interpolation with the expressive power of deep neural networks, our model is able to produce renderings where shadows and highlights move smoothly as a function of the light direction.

This work is by no means the final word for the task of light stage super-resolution or image-based rendering. Approaches similar to ours could be applied to other general light transport acquisition problems, to other physical scanning setups, or to other kinds of objects besides human subjects.
\changed{Though our network can work on inputs with different image resolutions, GPU memory has been a major bottleneck to apply our approach on images with much higher resolutions such as 4K resolution. A much more memory efficient approach for light-stage super-resolution is expected for production level usage in the visual effects industry.}
Though we exclusively pursue the one-light-at-a-time light stage scanning approach, alternative patterns where multiple lights are active simultaneously could be explored, which may enable a more sparse light stage design.
Though the undersampling of the light stage is self-evident in our visualizations, it may be interesting to develop a formal theory of this undersampling with respect to materials and camera resolution, so as to understand what degree of undersampling can be tolerated in the limit. We have made a first step in this direction with the graph in Fig.~\ref{fig:freq}.
% It is interesting to see if there is a way to develop a theory of
% appearance sampling, when considering learning-based superresolution
% methods.  This would point to the limits of undersampling which can be
% tolerated.
Finally, it would be interesting to extend our approach to
enable the synthesis of novel viewpoints in addition to lighting directions.
We believe that light stage super-resolution represents an exciting direction for future research, and has the potential to further decrease the time and resource constraints required for reproducing accurate high-frequency relighting effects.
% We believe that upsampling light transport data is an exciting direction,
% enabling reproduction of accurate high-frequency effects within the
% constraints of limited acquisition time and resources.